◘

\documentclass{jfm}
\usepackage{graphicx}
\usepackage{comment}
\usepackage{epstopdf, epsfig}
\usepackage{xcolor} 
\usepackage{enumitem}
\usepackage{amsmath}
\usepackage{mathrsfs}
\usepackage{amsmath,bm} 
\usepackage{multirow} 
\usepackage{booktabs} 
\usepackage{amsbsy} 
\usepackage{upmath} 
\usepackage{mathtools} 
\usepackage{amssymb} 
\usepackage{natbib} 
\usepackage{amsfonts} 
\usepackage{amsgen} 
\usepackage{bookmark}
\usepackage{textgreek}
\usepackage[utf8]{inputenc} 
\usepackage[T1]{fontenc}
\usepackage{CJKutf8}
\usepackage{hyperref}
\usepackage{lineno}

\newcommand{\AdjustBibOrder}[1]{}
\hypersetup{colorlinks, linkcolor=red, citecolor=black, urlcolor=green}
\makeatletter
\DeclareRobustCommand{\iscircle}{\mathord{\mathpalette\is@circle\relax}}
\newcommand\is@circle[2]{%
  \begingroup
  \sbox\z@{\raisebox{\depth}{$\m@th#1\bigcirc$}}%
  \sbox\tw@{$#1\square$}%
  \resizebox{!}{\ht\tw@}{\usebox{\z@}}%
  \endgroup
}

\usepackage{enumitem}
\newlist{myenumi}{description}{10}
\setlist[myenumi]{leftmargin=23pt,itemsep=2pt,topsep=2pt,parsep=2pt}
\newlist{myenumi2}{description}{10}
\setlist[myenumi2]{leftmargin=40pt,itemsep=2pt,topsep=2pt,parsep=2pt}

\makeatother

\graphicspath{{figure/}}
\shorttitle{Vortical structures in turbulent stratified shear layers}
\shortauthor{Jiang, Lefauve, Dalziel \&  Linden}
\title{The evolution of coherent vortical structures \\ \vspace{0.1cm}
in increasingly turbulent stratified shear layers}
\author{Xianyang ~Jiang\aff{1}\corresp{\email{xj254@cam.ac.uk}}, Adrien Lefauve\aff{1}, Stuart B. Dalziel\aff{1}
    \and P.~F.~Linden\aff{1}}
\affiliation{\aff{1}Department of Applied Mathematics and Theoretical Physics, Centre for Mathematical Sciences, University of Cambridge, Wilberforce Road, Cambridge CB3 0WA, UK}
\begin{document}
\maketitle

\begin{abstract}
\begin{CJK*}{UTF8}{gbsn}
We study the morphology of Eulerian vortical structures and their interaction with density interfaces in increasingly turbulent stably-stratified shear layers.  We analyse the three-dimensional, simultaneous velocity and density fields obtained in the stratified inclined duct laboratory experiment. We track, across 15 datasets, the evolution of coherent structures from pre-turbulent Holmboe waves, through intermittent  turbulence, to full turbulence and mixing. We use the  Rortex--Shear decomposition of the vorticity field into a pure rotational part (the rortex vector), and a non-rotational part (the shear vector).  
We describe the  morphology of ubiquitous hairpin-like vortical structures (revealed by the rortex), 
similar to those commonly observed in boundary-layer turbulence. These are born as relatively weak vortices around the strong three-dimensional shearing structures of confined Holmboe waves, and gradually strengthen and deform under increasing turbulence, transforming into pairs of upward- and downward-pointing hairpins propagating in opposite directions on the top and bottom edge of the shear  layer. 
Each hairpin's pair of legs are counter-rotating and entrain fluid laterally and vertically, and their arched-up `heads', which are transverse vortices, entrain fluid vertically. We then elucidate how this large-scale vortex morphology stirs and mixes the density field. Essentially, vortices located at the sharp density interface on either edge of the mixing layer (mostly hairpin heads) engulf blobs of unmixed fluid into the mixing layer, while vortices inside the mixing layer (mostly hairpin legs) further stir it, generating strong, small-scale shear, enhancing mixing. These findings provide new insights into the role of turbulent coherent structures in shear-driven stratified mixing. 
\end{CJK*}
\end{abstract}

\clearpage

\section{Introduction}
Stably-stratified turbulence and the enhanced mixing across density isosurfaces (isopycnals) that it accomplishes is a crucial but poorly-understood component of many deep-ocean and coastal flow systems of importance under a changing climate. A `grand challenge' of environmental fluid dynamics is to parameterise accurately this small-scale `diapycnal' mixing in large-scale circulation models to improve predictions for the vertical transport of heat, carbon dioxide, salt, and other scalars in our oceans  \citep{dauxois_confronting_2021}.  To complement expensive and sparse field observations, laboratory experiments have a key role to play in the effort to develop better mixing parameterisations. In this paper we will use datasets obtained from such an experiment, the stratified inclined duct (SID), which sustains a two-layer exchange flow in an inclined square duct. This experiment allows us to accurately control the flow geometry and levels of interfacial turbulence by a systematic variation of two key non-dimensional flow parameters. Using newly-available measurement technologies \citep{partridge2019}, this experiment also allows us to obtain the three-dimensional Eulerian velocity and density fields simultaneously at high spatio-temporal resolutions, and thus to study  three-dimensional coherent structures like never before. 

These coherent flow structures -- and especially vortical structures -- exist across a wide spectrum of spatio-temporal scales and play an important role in the processes of turbulent bursting and mixing.  Previous studies identified a range of vortical structures in stratified shear layers (i.e. in a nearly-parallel layer of vorticity not caused by a solid wall, and that embeds a density interface). Among them are: streamwise or quasi-streamwise vortices \citep{Schowalter1994,Caulfield2000}, hairpin vortices \citep{Smyth2003,watanabe2019}, spanwise vortices \citep{Salehipour2015}, columnar vortices \citep{Billant2000,Waite2008},  and pancake vortices \citep{Fincham1996,Riley2000,Riley2003}. Most of these  were identified by Eulerian criteria, either based on a threshold of vorticity \citep{Basak2006,Mashayek2012} or through the eigenvalues of the velocity gradient tensor \citep{Hunt1988,Chong1990,Jeong1995,Zhou1999,Chakraborty2005}. In the shear-driven flows of interest in this paper, meaningful turbulent vortical structures must be defined after an appropriate treatment of the `contaminating' mean shear \citep{Shrestha2021}. Recent efforts have been devoted to decompose the velocity gradient tensor into a rotational and shearing part \citep{Li2014,Keylock2018,Gao2018,Nagata2020,Watanabe2020,Hayashi2021}. Here, we apply to our state-of-the-art experimental datasets the new Rortex--Shear (RS) decomposition proposed by \cite{Wang2019} and \cite{Xu2019} to decompose the three-dimensional (3-D) vorticity field into a `pure rotation' field and a `pure shear' field. 

Although vortices (i.e. coherent regions of pure rotation) can  be produced \emph{externally} (e.g. by artificial vortex rings impinging  a density interface as a model for turbulent eddies, see \cite{Linden1973,Olsthoorn2015}), they naturally develop \emph{internally}, either from internal gravity waves \citep{Fritts1998} or, more typically, from shear-driven instabilities leading to (usually short-lived) Kelvin--Helmholtz billows \citep{Caulfield2000} or (usually long-lived) Holmboe waves \citep{Smyth2003}. \citet{Lefauve2018} described and explained the origin of `confined' Holmboe waves in the SID experiment,  a typical example of long-lived coherent vortical structures, which they visualised using a simple vorticity threshold. The 3-D development of the Holmboe-wave instability was studied numerically by \cite{Smyth2003}, who noted that ``Loop structures in the density field associated with hairpin-like vortices are a conspicuous feature of turbulent Holmboe waves. These structures are initiated by secondary instabilities (in one case this resembled the localised convective instability described by \cite{Smyth1991}) and grow to large amplitude via vortex stretching.'' The hypothesis of horseshoe (or hairpin) vortices was initially proposed by \cite{Theodorsen1952}, and has proven key to the understanding of boundary-layer turbulence \citep{Acarlar1987_2,Smith1991,Adrian2007,Jiang2019AMM,Lee2019}. \cite{Head1981} used smoke visualisation to investigate the evolution of hairpins in turbulent boundary layer with increasing Reynolds number,  and they found that elongated hairpin vortices were inclined at a characteristic angle of approximately $40-50^\circ$ to the wall. These hairpins were observed to be less elongated and  more isolated at low Reynolds number, and to agglomerate and become very elongated at high Reynolds number. The inclination and evolution (generation and regeneration) of hairpins was subsequently studied in more details, numerically by \cite{Zhou1999} and experimentally by \cite{Haidari_Smith1994}. A hypothesis based on soliton-like coherent structures has been put forward to explain the bursting process and the generation of hairpins in wall-bounded flows \citep{Lee1998,Lee2008,Jiang2020,Jiang2020-2}. Hairpin-like structures have also been observed in stably-stratified boundary layers, experimentally by \cite{Williams2014} and numerically by \cite{Atoufi2019}; they are apparently similar to those found in unstratified boundary layers. 

In stably-stratified shear layers (not visibly influenced by top and/or bottom walls), such as deep ocean overflows, exchange flows through straits, or saltwater intrusions in estuaries, Kelvin--Helmholtz or Holmboe instabilities (found in weakly- and strongly-stratified flows, respectively) can grow in a symmetric or asymmetric fashion depending on the vertical offset between the centres of the velocity profile and the density profile \citep{Carpenter2007}. Hairpin vortices have been associated with those wave trains, especially after they succumb to secondary instabilities (i.e. further symmetry breaking in the third dimension), whose breakdown creates fully 3-D turbulence \citep{Smyth2000,Smyth2006,Pham2012}. Recently, using direct numerical simulations, \cite{watanabe2019} found inclined hairpin vortices throughout the stratified shear layer, and argued that turbulent mixing was very active at the length scales close the streamwise extent of the hairpins. In stratified plane Poiseuille flow, \cite{Lloyd2022}  found numerically that hairpin vortices arise far from the wall and interact with strong buoyancy gradient, inducing internal wave breaking. 

However, despite tantalising numerical evidence of their existence and their important role in stratified shear-driven mixing, hairpin vortices have until now not been described in comparable laboratory flows. It also remains unclear (i) how they develop from pre-turbulent flows (especially Holmboe waves) and evolve in increasingly turbulent flows; and (ii) how they interact with density interfaces and participate in density overturning, stirring, and ultimately mixing. These are the two questions that we will address in this paper.

In \S~\ref{sec:datasets} we introduce our experimental datasets and explain their relevance to our objectives. In \S~\ref{sec:Identification} we visualise vortical structures first by a traditional method, and then by our new method based on the Rortex--Shear decomposition of vorticity, to build intuition for the subsequent statistical analyses. In \S~\ref{sec:anatomy} we reveal the detailed morphology of the `rortices' identified by the rortex (and, to a lesser extent, of the shear) by a `weighted conditional averaging' method. In \S~\ref{sec:interaction} we study the interaction between rortices and density gradients. In \S~\ref{sec:discussion} we synthesise these results and propose a tentative model for the origin and role of hairpin vortices in stratified shear layers, and we conclude in \S~\ref{conclusion}.

\section{Experimental datasets} \label{sec:datasets}
\subsection{Setup and flows}
The datasets analysed in this paper were collected in the Stratified Inclined Duct (SID) experiment, whose setup is described in prior publications such as \cite{Lefauve2018} (see their \S~3). The SID sustains a hydraulically-controlled exchange flow inside a long duct (of length $L=1350$~mm) of square cross-section (of height and width $H=45$~mm). The duct is inclined at a small angle $\theta$ with respect to the horizontal and connects two large reservoirs initially filled with aqueous salt solutions of different densities $\rho_0 \pm \Delta \rho/2$. The Prandtl number is thus $Pr=\nu/\kappa\approx 700$, where $\nu$ and $\kappa$ are the average kinematic viscosity and salt diffusivities, respectively.  

Increasing  $\theta$ (defined to be positive when it accelerates the flow) and/or the Reynolds number $Re \propto \sqrt{g(\Delta\rho/\rho_0)H}H/\nu$ (primarily set by the density difference) allows the experimenter to sweep through four qualitatively different flow regimes: from laminar flow with a flat interface, through finite-amplitude Holmboe waves propagating at the interface (this regime is hereafter abbreviated `H') then intermittent turbulence and interfacial mixing (hereafter `I') and to fully-developed turbulence and  mixing (hereafter `T'). These flow regimes  have been mapped in the ($\theta,Re$) plane and their transitions have been studied extensively  \citep{meyer2014,Lefauve2019,lefauve2020}. 

\subsection{Measurements and processing}\label{sec:datasets-processing}

We consider 15 datasets, each corresponding to a single experiment performed at a given $\theta$ and $Re$. Four belong to the H regime (labelled H1-H4), eight to the I regime (I1-I8), and three to the T regime (T1-T3). Each dataset comprises a time-resolved series of the three-component velocity field $(u,v,w)$ and density field $\rho$ given simultaneously in 3-D volumes $(x,y,z)$, where $u$ and $x$ are the streamwise velocity and coordinate (along the duct), $v$ and $y$ are spanwise, and $w$ and $z$ are `vertical' (normal to both $x$ and $y$) in the frame of reference of the tilted duct. The acceleration of gravity $\bm{g}$ along the `true vertical' is thus tilted with respect to this coordinate system and has components $[g\sin\theta,0,-g\cos \theta]$ along $(x,y,z)$. See \cite{Lefauve2019}, figure 1 for a schematic and figures 3--4 for a snapshot of $u$ and $\rho$ in each regime. 

These 3-D volumes were obtained by the novel laser-sheet-scanning technique described in \cite{partridge2019},  in which simultaneous stereo particle image velocimetry (PIV) and planar laser induced fluorescence (PLIF) are performed in  successive $x-z$ planes. The $u,v,w,\rho$ data obtained at spanwise locations $y = y_i$ ($i= 1, 2, \ldots , n_y$) and respective times $t = t_i$ are subsequently combined in volumes containing $n_y$ planes spanning the cross-section of the duct. This makes the volumes only `near-instantaneous' in the sense that each plane $(x, y_i, z, t_i)$ is separated from the previous one by a small time increment $\delta t = t_i - t_{i-1}$. The advantage of this method over earlier scanning or tomographic methods is the ability to scan relatively large volumes (here typically $200 \times 45 \times 45$~mm${}^3$) and obtain high $x-z$ planar resolutions for both velocity and density. {Each experiment typically captures $\approx 300$ volumes (time snapshots), and each volume typically contains  $\approx 400\times 40\times80$ velocity vectors in $x,y,z$, respectively.}

Instead of the original datasets used and visualised in \cite{Lefauve2018,Lefauve2019,partridge2019},  in this paper we use the slightly modified datasets of \cite{Lefauve20221,Lefauve20222} (hereafter LL22a,b). These modifications are explained in LL22a (see their \S\S~3.3-3.5 and figure 1) and are summarised as follows. First, they cropped early transients (in $t$) characterised by a slight net flow through the duct (sloshing between reservoirs) in order to focus on statistically-steady dynamics. Second, they cropped the near-wall regions (in $y$ and $z$) in order to discard viscous boundary layers and focus on the interfacial quasi-hyperbolic-tangent `free shear layer' region. Third, they non-dimensionalised the coordinates and flow variables of each individual dataset using (i) half the density difference $\Delta \rho/2$ (after removing the mean $\rho_0$) such that $-1\le\rho\le 1$; (ii) half the actual resulting shear-layer depth, such that $-1\le z \le 1$; and (iii) half the actual (mean) peak-to-peak velocity magnitude, such that $-1\lesssim  u \lesssim 1$ {with} the mean velocity extrema  $\langle u \rangle_{x,y} (y=0,z=\pm 1) = \mp 1$. This {cropping procedure} allows for a  meaningful side-by-side non-dimensional analysis of the shear-layer dynamics of all 15 datasets.

The datasets, and the associated codes and 3-D visualisation movies (with the same viewing angle and slices positions as in this paper), can all be freely downloaded from their repository \cite{lefauve2022dataset} .

\subsection{Parameters and resolution}

The corresponding `shear-layer' Reynolds number $Re$  and bulk Richardson number $Ri_b$ are defined as in LL22a \S~3.2-3.3 as %
\begin{equation} \label{def-Re}
    Re \equiv \underbrace{\frac{\dfrac{
    \Delta U}{2}\dfrac{H}{2}}{\nu}}_{\begin{subarray}{c}\text{hydraulics}\\
    \text{(input)}\end{subarray}} \, \cdot \,
    \underbrace{\dfrac{\delta u}{2}\dfrac{h}{2}}_{\begin{subarray}{c}\text{shear layer}\\
    \text{(output)}\end{subarray}} \equiv \frac{\sqrt{g'H}H}{2\nu}\cdot  \frac{\delta u \, h }{4}
\end{equation}
and
\begin{equation} \label{def-Rib}
 Ri_b \equiv \underbrace{\frac{\dfrac{g}{\rho_0}\dfrac{\Delta \rho}{2}\dfrac{H}{2}}{\Big(\dfrac{\Delta U}{2}\Big)^2}}_{\begin{subarray}{c}\text{hydraulics}\\
    \text{(input)}\end{subarray}}  \, \cdot \, \underbrace{\dfrac{\dfrac{h}{2}}{\Big(\dfrac{\delta u}{2}\Big)^2}}_{\begin{subarray}{c}\text{shear layer}\\
    \text{(output)}\end{subarray}} \equiv \frac{1}{4} \cdot \frac{ 2h }{(\delta u )^2}.
\end{equation}
These parameters are consistent with the  non-dimensionalisation introduced above. These parameters consist of (i) a `hydraulics' part based on input parameters, including half the peak-to-peak dimensional velocity scale $\Delta U/2 \equiv \sqrt{g'H} = \sqrt{g(\Delta \rho/\rho_0)H}$, half the duct height $H/2$ and $\nu$, and (ii) a `shear-layer' rescaling based on half the non-dimensional output (measured after the hydraulic non-dimensionalisation) peak-to-peak velocity magnitude $\delta u/2$ (typically $\approx 0.5-1.2$) and shear layer depth $h/2$ (typically $\approx 0.5-0.7$).
Note that $Re$ and $Ri_b$ were denoted as $Re^s$ and $Ri_b^s$, respectively, in LL22a,b to emphasise this specific shear-layer non-dimensionalisation. 

The parameters of datasets H1-T3 are shown in  table~\ref{tab:dataset}. 
For further properties, such as the mean flows, see LL22a, \S~4. As a rule of thumb, increasing levels of turbulence and transitions between the H, I, and T regimes, are well described by the product $\theta  Re$. The historical names of datasets  (H1, $\ldots$ H4, I1, $\ldots$ I8, T1, $\ldots$, T3) were based on increasing values of the product of $\theta$ with the hydraulics (input) Reynolds number $\Delta U H/(4\nu$) in  \cite{Lefauve2019} rather than on the shear-layer (output) Reynolds number $Re$ used in this paper. Nevertheless, datasets remain approximately ordered with increasing $\theta  Re$.

\begin{table}
  \begin{center}
\def~{\hphantom{0}}
\setlength{\tabcolsep}{4pt}
  \begin{tabular}{c|ccccccccccccccc}
Name & H1 & H2 & H3 & H4 & I1 & I2 & I3 & I4 & I5 & I6 & I7 & I8 & T1 & T2 & T3 \\[3pt]
$\theta$ ($^\circ)$    &  1 &   5   &  2  &  5 &    2  &   2   &  2   &  6   &  5  &   6  &   3   &  5   &  3    & 6   &  5 \\[5pt]
$Re$      &   381   &      204   &      422      &   203      &   531     &    872  &       891     &   646      &   607     &    497   &      905     &    708    &    1479  &      1030     &   1145 \\ [5pt]
$\theta Re$    &  7  & 18  & 15 &  18  & 19  & 30  & 31 & 68 &   53 &  52 &  47 &  62  & 77 & 108  & 100
  \end{tabular}
  \caption{List  of the 15 volumetric data sets used, adapted from LL22a's table 1. Note that $\theta$ is in radians in the product $\theta Re$.  }
\label{tab:dataset}
  \end{center}
\end{table}

As a rule of thumb, the vector resolution of the data approaches the Kolmogorov turbulent lengthscale (marking the end of the inertial range)  in $x,z$, but it is coarser in $y$. The temporal resolution (i.e. the time it takes to reconstruct a single volume) is of order 1--4 advective time units, smaller values representing a better `freezing' of the flow. The full vector resolution of each dataset in $x,y,z,t$ are given in LL22a, table~3 (datasets with the best spatial and  temporal resolutions are highlighted in bold).

Although subject to inherent technical limitations  (summarised in \cite{Lefauve2019} Appendix~A, and LL22b Appendix B), we will show below that these 15 state-of-the-art datasets deliver new insights into the time-resolved, 3-D coherent structures of shear-driven stratified turbulence.

\section{Identification of vortical structures}
\label{sec:Identification}

\subsection{Previous methods and $Q$-criterion}

Since this paper focuses on Eulerian 
vortical structures in shear layers, we start by addressing the delicate first step of identifying such `vortices'.   It is known that identifying a vortex based on a specified threshold of the magnitude of the vorticity vector $\bm{\omega}=\boldsymbol{\nabla}\times \bm{v}$ is subjective and generally inappropriate \citep{Zhou1999,Gao2011}. For example, $\bm{\omega}$ is not generally aligned with the local rotation, and the maximum of $|\bm{\omega}|$ does not generally coincide with the `core' of a vortex, since vorticity does not discriminate between shear and swirl (rotation). 

Several improved Eulerian vortex identification schemes based on the eigenvalues of the velocity gradient tensor {$\boldsymbol\nabla\bm{v}$} have thus been developed since the 1980s, including the $Q$-criterion \citep{Hunt1988}, $\Delta$-criterion \citep{Chong1990}, $\lambda_2$-criterion \citep{Jeong1995}, $\lambda_{ci}$-criterion \citep{Zhou1999,Chakraborty2005}, etc. These methods have proved effective and influential to study approximate vortex boundaries in a variety of flows. 

\begin{figure}
\centering
\includegraphics[width=0.94\textwidth]{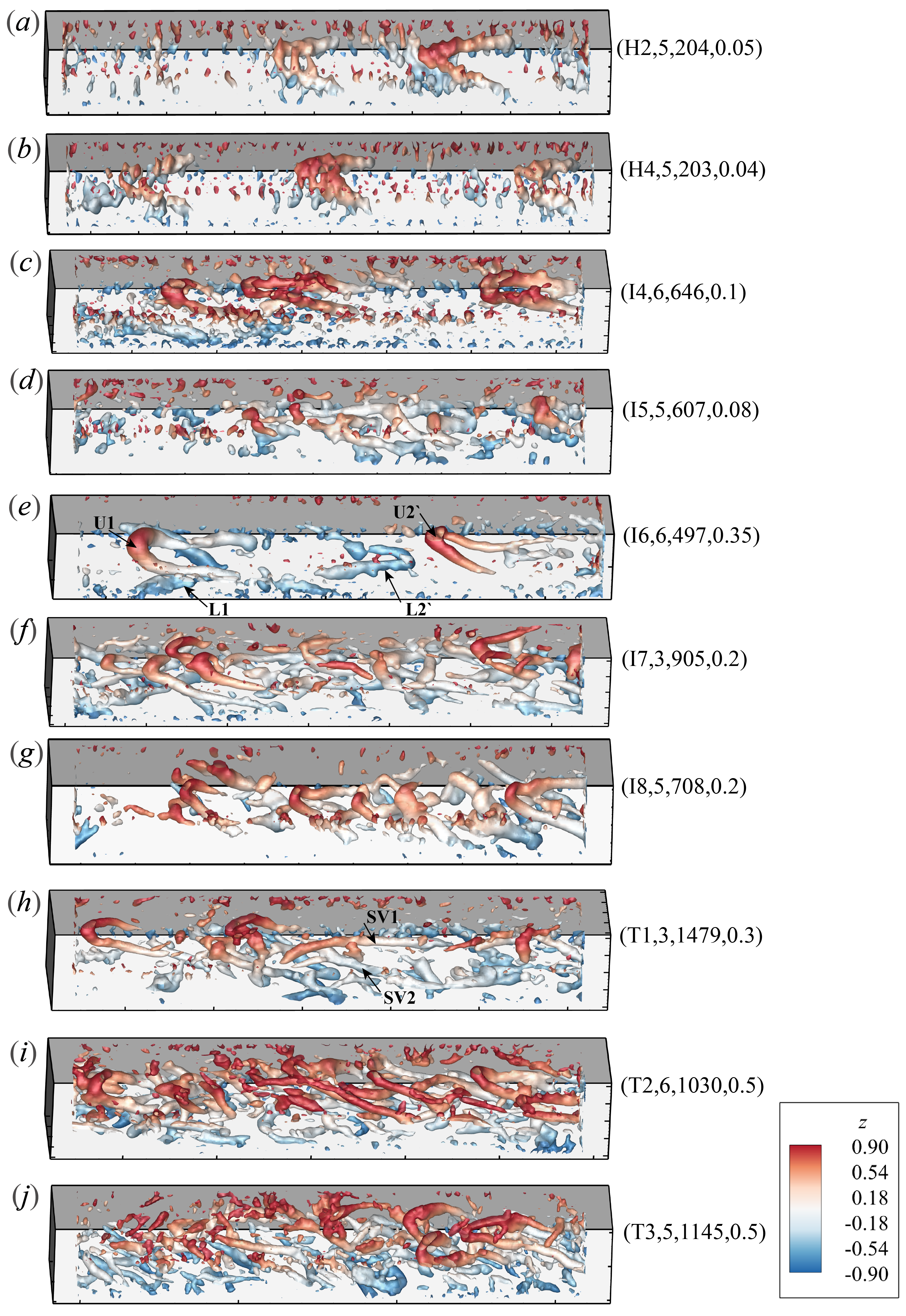}
\caption{Isosurfaces of the $Q$-criterion (single snapshot) in datasets  (\textit{a}) H2, (\textit{b}) H4, (\textit{c}) I4, (\textit{d}) I5, (\textit{e}) I6, (\textit{f}) I7, (\textit{g}) I8, (\textit{h}) T1, (\textit{i}) T2 and (\textit{j}) T3. The dataset name, tilt angle $\theta$, shear-layer Reynolds number $\Rey$ (see table~\ref{tab:dataset}) and the iso-surface $Q$ value are listed on the right of each panel. Colours on the iso-surfaces denotes the $z$ position (we show $-0.9< z<0.9$, i.e. the middle 90\% of the shear layer). }
\label{fig:Q}
\end{figure}

Thus we begin our analysis of vortical structures in figure~\ref{fig:Q} by a visualisation of vortices based on the popular $Q$-criterion, {i.e.} the second principal invariant of $\boldsymbol{\nabla}\bm{v}$ calculated as $Q=||\bm B||^2-||\bm A||^2$, where $\bm A =\frac{1}{2}(\boldsymbol\nabla\bm{v} + (\boldsymbol\nabla\bm{v})^T)$ and $\bm B =\frac{1}{2}(\boldsymbol\nabla\bm{v} - (\boldsymbol\nabla\bm{v})^T)$ are, respectively, the symmetric (or strain rate tensor) and anti-symmetric (or rotation rate tensor) components of the $\boldsymbol{\nabla}\bm{v}$. We plot a single snapshot of $Q$-criterion vortices identified by the isosurface $Q>0$ (i.e. the local rotation exceeds strain) in a selection of 10 representative datasets having three different tilt angles: $\theta = 5^\circ$ in panels (\textit{a,b,d,g,j}) (datasets H2, H4, I5, I8, T3, with increasing $\Rey$), $\theta = 6^\circ$ in panels (\textit{c,e,i}) (datasets I4, I6, T2) and $\theta = 3^\circ$ in panels (\textit{f,h}) (datasets I7, T1). The colouring of the iso-surfaces denotes the vertical location $z$ and the legend on each plot identify the dataset, the tilt angle, the Reynolds number {and the $Q$ value of the iso-surface displayed}.

Broadly speaking, from the Holmboe to the turbulent regime (i.e. with increasing $\theta \Rey$), vortical structures evolve from individual, disconnected hairpins which  start as \textLambda-structures without elongated trailing legs (panels~\emph{a,b}), to groups of hairpins with elongated legs  (panels~\emph{h,i,j}). The hairpins of low-$\Rey$ flows are relatively weak and less obvious (see panels~\emph{c,d}), especially when a corrugation appears on the iso-surfaces (which we attribute to a low signal-to-noise ratio), or when the head of the hairpin is broken. However, we will show below that these hairpin-shaped vortices are indeed robust features of Holmboe waves. 
In higher-$\Rey$ flows, hairpins are stronger and more obvious (higher signal-to-noise ratio), with their head tending toward the edges of the shear layer ($|z|\approx 1$) and their legs stretching in the streamwise direction ($x$).

Further features are worth mentioning. Figure~\ref{fig:Q}(\textit e)  shows two large hairpins in each of the upper (isosurfaces shaded in red and labelled `U1' and `U2') and lower (shaded in blue and labelled `L1' and `L2') parts of the shear layer.  Panels (\textit{f,g,h}) (weaker turbulence) show hairpins that are usually asymmetric in the sense that one `leg' is longer than the other. In this paper we interpret `quasi-streamwise vortices' ({denoted by `SV'; e.g. see} arrows SV1, SV2 in panels \textit{h}) as an extreme form of asymmetric hairpin vortices. Panels~(\textit{i,j}) (the most turbulent datasets) show  large-scale hairpins coexisting with  small-scale, fragmentised structures distributed over the shear layer, which form hairpin packets/forests reminiscent of turbulent boundary-layers. 

The supplementary movie 1 shows the complete time evolution of panels (\textit e-\textit j). 
From these snapshots (and supplementary movie 1), we hypothesise that hairpin-like coherent vortical structures may be a common and important vortical structure in SID flows.
\subsection{Rortex--Shear decomposition}

The  scalar $Q$-criterion (i.e. the isosurface  $Q=Q_0$, where $Q_0$ is the threshold) is inevitably subjective to the (somewhat arbitrary) value of the threshold. Furthermore, Q-vortices do not represent rigid rotation since they are contaminated by shear \citep{Liu2019}.

Recently, a new `vortex' vector called the `rortex' (or `liutex') was proposed by \cite{Liu2018} and \cite{Gao2018} to isolate the rigid rotational part from the shear, and directly point in the direction of local fluid rotation. The vorticity is decomposed uniquely as $\bm{\omega} = \bm{R} + \bm{S}$, where  $\bm{R}$ is the rortex vector, and $\bm{S}$ is the shear vector (indicating a non-rotational, anti-symmetric part of $\boldsymbol\nabla\bm{v}$). This Rortex--Shear (RS) decomposition thus separates pure rotational structures from shearing structures. This decomposition is particularly helpful in shear-driven turbulence, as in this paper. Furthermore, \cite{Xu2019} showed that the RS decomposition was related to a special (transposed) Schur form of the velocity gradient tensor (rotating the $\boldsymbol\nabla\bm{v}$ into a suitable block-triangular matrix). This relationship allowed the use of efficient library functions to  speed up the RS decomposition by avoiding the tedious original algorithms requiring coordinate transformations.

According to \cite{Xu2019} and \cite{Wang2019}, the rortex  could then be explicitly calculated as 
\begin{equation}
    {\bm{R}=\left(1-\sqrt{1-\frac{4\lambda_{ci}^2}{(\boldsymbol\omega\cdot\bm{r})^2}}\right)(\boldsymbol\omega\cdot\bm{r})\bm{r}}.
    \label{RotexEqn}
\end{equation}
The direction of the rortex $\bm{r}$ is the local unit \emph{real} eigenvector of   $\boldsymbol{\nabla} \emph{\textbf{v}}$, indicating the rotational axis, and $\lambda_{ci}$ is the imaginary part of the complex conjugate part of eigenvalues of $\boldsymbol{\nabla} \emph{\textbf{v}}$.
This is the practical definition that we apply to our datasets in the remainder of this paper. It is based on the idea that $\boldsymbol\nabla\bm{v}$ can have either one or three real eigenvalues. When there is only one real eigenvalue, the direction of the rortex $\bm{R}$ is aligned with the associated normalised eigenvector $\bm{r}$ selected such that {$\boldsymbol\omega\cdot\bm{r} > 0$}. The conjugate pair of complex eigenvalues have imaginary parts $\pm\lambda_{ci}$ characterising the rotation about $\bm{r}$. When there are three real eigenvalues, $\lambda_{ci} = 0$ and thus $\bm{R}=\bm{0}$, meaning that all the vorticity is shear without rotation. Equation \ref{RotexEqn} shows the relative importance of $\lambda_{ci}$ \emph{vs} the vorticity projected onto $\bm{r}$. 

Hereafter, we denote the magnitude of the shear vector $|\bm{S}|=S$, and we shall refer to the magnitude of the rortex vector $|\bm{R}|=R$ as the \emph{rorticity} and the vortical structure it identifies simply as a \emph{rortex}. We shall also use the term \emph{vortex} to refer more generically to vortical structures that have not been unequivocally identified using the RS decomposition, as is the case in all the literature pre-dating 2018.

\subsection{Rortex and shear structures}

\begin{figure}
\centering
\hspace{-0.3cm}
\includegraphics[width=\textwidth]{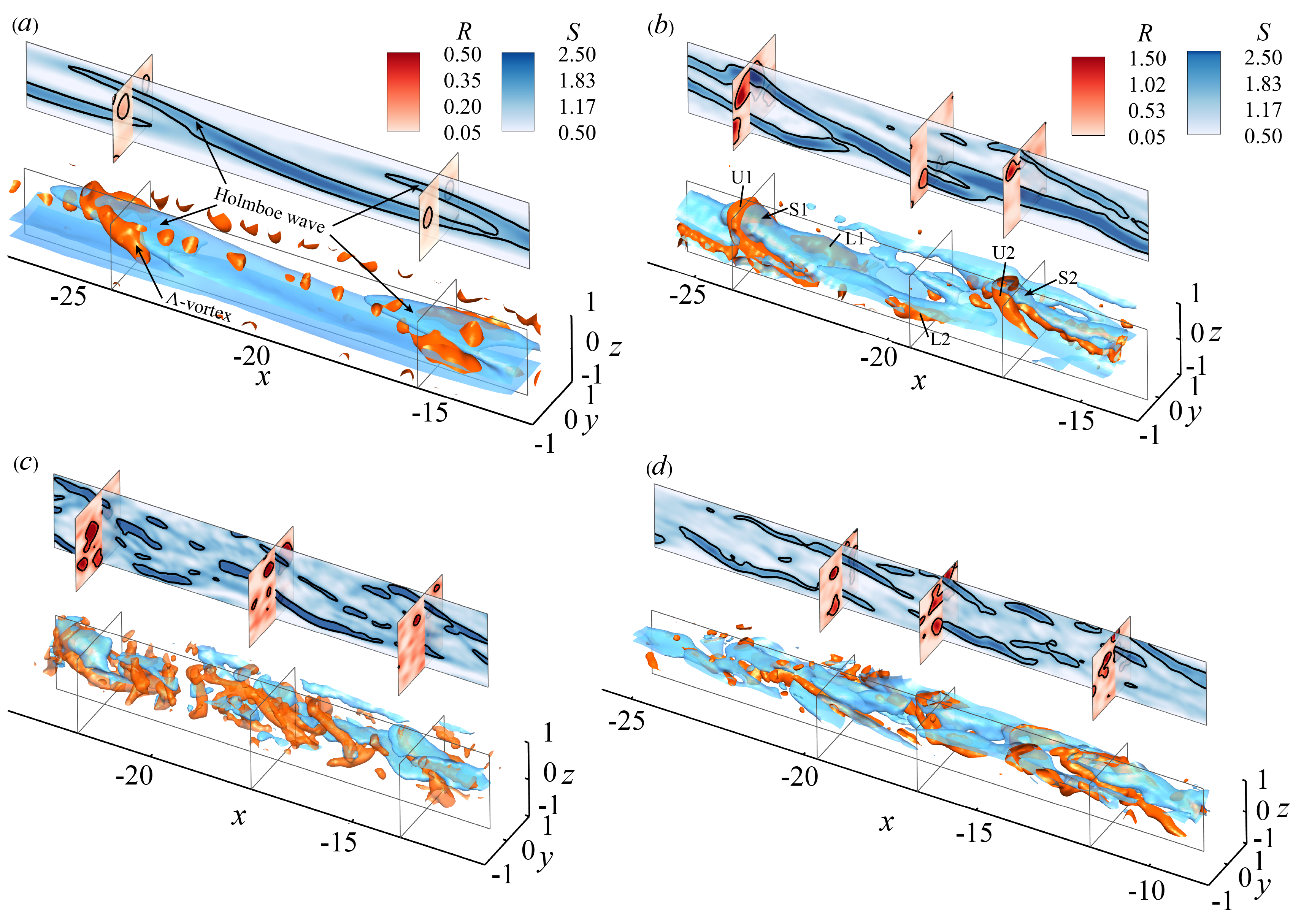}
\caption{Visualisation of rortex ($R$) and  shear ($S$) structures in datasets (\textit a) H4  (snapshot at time $t_n=180$), (\textit b) I6  (at $t_n=36$); (\textit c) T2 (at $t_n=39$) and (\textit d) T3  (at $t_n=132$). In each panel, $y$-$z$ slices show the $R$ values, while the $x$-$z$ planes show the $S$ values. Red 3-D structures represent an $R$ iso-surface (respectively $R=0.13, \, 0.5, \, 0.85, \, 0.8$ in \textit{a, b, c, d}), while blue structures represent an $S$ iso-surface (respectively $S=1.3, \, 1.4, \, 2, \, 1.8$ in \textit{a, b, c, d}). The value of the iso-surfaces are shown by a black contour on the respective slices. Note that (\emph{c-d}) share the same colour scales for $R$ and $S$ as panel (\textit b). In all isosurface panels, the spanwise edges of the shear layer $|y|>0.85$ have been blanked for clarity, so has been the lower half $z<0$ in (\emph{c-d}).
}
\label{fig:RSh4I6T2T3}
\end{figure}

Figure \ref{fig:RSh4I6T2T3} shows an iso-surface of both rortex $R$ (in red) and shear $S$ (in blue) in four datasets (H4, I6, T2 and T3), together with  contour plots of $R$ in $y-z$ planes (cross-sectional slices, red shades) and a contour plot of $S$ in an $x-z$ plane (longitudinal slice, blue shades). Black contour lines highlight the value of each iso-surface and its projection on the respective slices. (See the supplementary movies 2, 3, 4 and 5 for the complete evolution of $R$ and $S$ as well as the velocity and density information around them for Holmboe, intermittency and turbulent regimes, respectively.) 

Figure \ref{fig:RSh4I6T2T3}(\textit a) in the Holmboe regime (H4) shows a pure shearing structure (blue iso-surface), which is highly reminiscent of the shape of the confined Holmboe wave described in  \cite{Lefauve2018} from iso-surfaces of the spanwise component of vorticity ($\omega_y$) of the same dataset. This similarity is because the shear is very strong in the Holmboe regime, as seen by the fact that contour values for $S$ are at least five times larger than that of $R$ (see colour bars), and that the blue isosurface $S=1.3$ is ten times larger than the red isosurface $R=0.13$. This indicates that shear accounts for a substantial part of the vorticity according to $\bm\omega=\bm R+\bm S$. Although weaker, rortices are also observed near the `head' of the Holmboe wave, in a \textLambda-shape similar to the $Q$-vortex from figure \ref{fig:Q}(\textit{a-b}). The two `legs' of the rortex flank the `head' of Holmboe wave, leading to a hypothesis that the rortex may originate from the localised high shear regions of the wave.

Figure \ref{fig:RSh4I6T2T3}(\textit b) in the  low-$Re$ intermittent regime (I6) shows slightly evolving $R$ and $S$ patterns, with a hairpin rortex straddling the shear, as pointed out by the arrows S1 and S2. The snapshot in figure \ref{fig:RSh4I6T2T3}(\textit b) is {for the same flow and time as} that shown in figure \ref{fig:Q}(\textit e), where two pairs of upper and lower rortices were labelled U1, U2, L1 and L2, respectively. These $S$ structures in figure \ref{fig:RSh4I6T2T3}(\textit b) seem to originate from increasingly turbulent \emph{symmetric} Holmboe waves (having upper and lower counter-propagating modes), as opposed to the \emph{asymmetric Holmboe wave} of figure  \ref{fig:RSh4I6T2T3}(\textit a) (only having an upper mode).

Figure \ref{fig:RSh4I6T2T3}(\textit{c,d}) in the turbulent regime (T2 and T3) shows more numerous and smaller-scale structures, such that the lower half of the shear layer ($z<0$) was {omitted} for clarity.  An apparently robust observation is that rortices still tend to straddle the strong shearing region. The high-shear regions tend to be pushed nearer the top and bottom edges of shear layer, and they are more aligned with the $x$ direction (less tilted) than in the Holmboe and intermittent regimes. Finally, the relative strength of rorticity \emph{vs} shear also increases (see the colour bars and isosurface values in the caption), indicating an increasing importance of rortices in  turbulent mixing.

\subsection{Averaged  distribution of rorticity and shear}

\begin{figure}
\centering
\makebox[\textwidth][c]{\includegraphics[width=\textwidth]{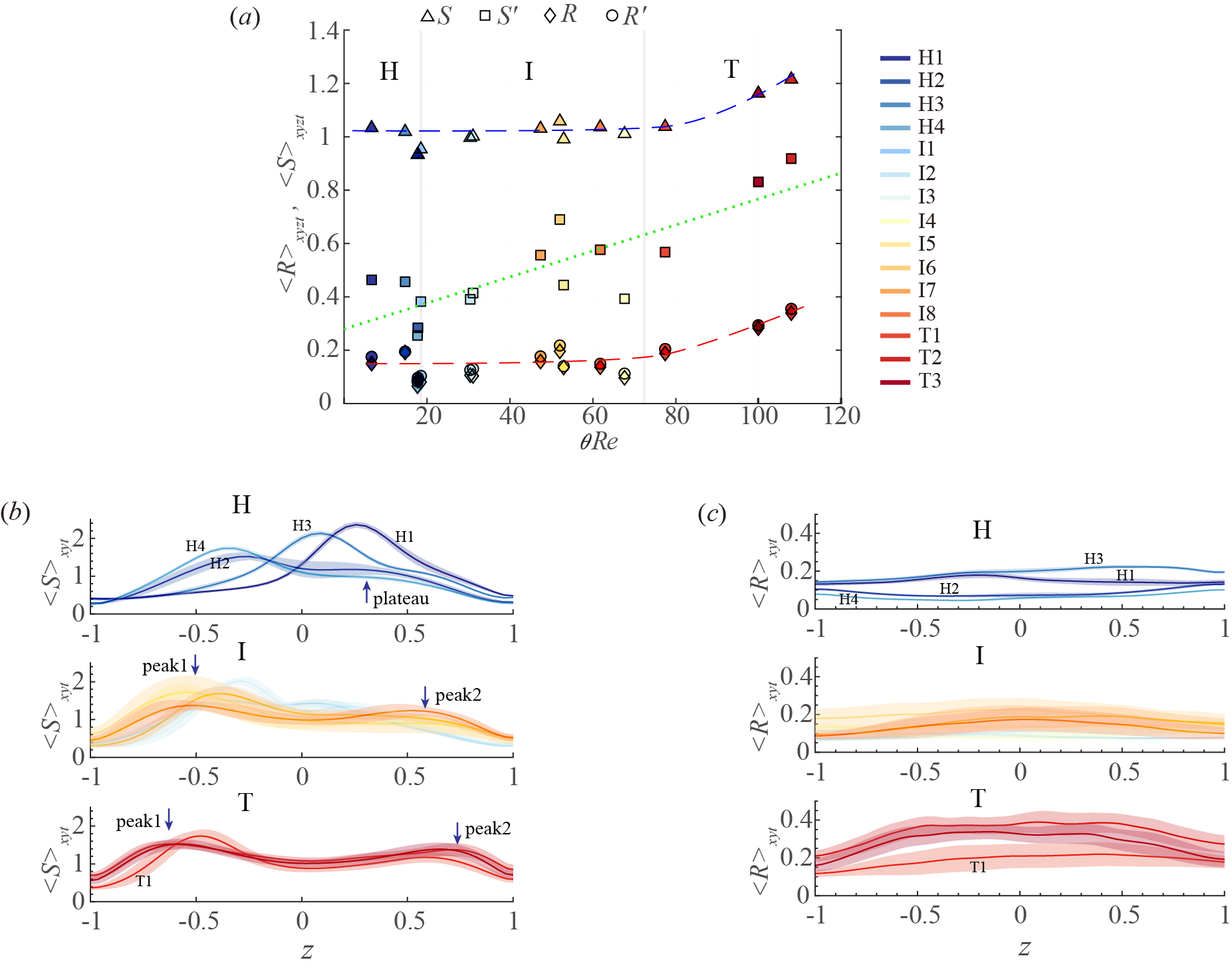}}
\caption{(\textit{a}) Distribution of $\langle  R\rangle_{xyzt}$ and $\langle  S\rangle_{xyzt}$ for all datasets, {separating H, I, and T data by the solid grey lines} ($\triangle$ for $S$;  $\diamondsuit$ for $R$; $\square$ for $S'$; $\iscircle$ for $R'$; {the dashed lines are trend lines of the distribution;} the dotted line is {the} fitting curve for $S'$; {filled colours of the symbols denote the flow regime and number shown on the right}).  (\textit{b}) and (\textit{c})  are distribution of $\langle  S\rangle_{xyt}$ and $\langle  R\rangle_{xyt}$ along $z$ direction for all datasets, respectively, separating H, I, and T data in different rows. {Colours of the curves and shadings share the same legend in (\textit a)}. The transparent shadings denote the local variability in time corresponding to one root-mean-square value of the $S$ and $R$ (i.e. $\langle S\rangle_{xy}-\langle S\rangle_{xyt}$ and similarly for $R$).}
\label{fig:RS_SL}
\end{figure}

The  relative strengths of $R$ and $S$ are explored quantitatively in figure \ref{fig:RS_SL}. In panel (\textit{a}) we plot, for all 15 experimental datasets, the averaged magnitudes $\langle R\rangle_{xyzt}$ and $\langle S\rangle_{xyzt}$ (where $\langle \cdot\rangle_{xyzt}$  represents the average over the whole shear layer volume and time as in LL22a,b) against the product $\theta Re$ (our proxy for increasing levels of turbulence). Both $\langle R\rangle_{xyzt}$ ($\diamondsuit$) and $\langle S\rangle_{xyzt}$ ($\triangle$) are nearly constant at $\approx 0.1-0.2$ and $\approx 1$, respectively, when $\theta \Rey\lesssim 80$ (where $\theta$ is in radians), corresponding to the H and I regimes, but they increase with turbulence beyond this (see the dashed trend lines).

To understand this, we also performed the RS decomposition on the fluctuating velocity  $\bm v'=\bm v -  \bar{\bm v}$ where $\bar{\cdot} = \langle \cdot \rangle_{t}$ is the {spatially varying} temporal average, giving the underlying parallel shear flow $\bar{u}(x,y,z)$ ({since} $\bar{v},\bar{w}\approx 0$). Figure  \ref{fig:RS_SL}(\textit a) also shows the resulting fluctuating shear $S'$ ($\square$) and rortex  $R'$ ($\iscircle$). First, we find that $R \approx R'$ (the symbols are nearly indistinguishable) whereas $S' \ll S$, as expected in the {presence} of background shear $\partial_z \bar{u}$. Second, we find that $S'$ is only about two to three times larger than $R'$ (a weaker dominance compared to that of $S$ over $R$), and $S'$ seems to increase fairly linearly with $\theta \Rey$ even before the {transition to fully turbulent flow} at $\theta \Rey \approx 80$ (see the green dotted trend line).
These observations suggest that the background shear  plays an important role, but the details are beyond the scope of the present study, which focuses primarily on rortex structures. The remainder of this paper thus adopts the RS decomposition of the total velocity, as in the original papers of \cite{Liu2018} and \cite{Gao2018}.

To examine the strength of $S$ and $R$ along the $z$ (vertical) direction, we plot $x,y,t$ averages in figure  \ref{fig:RS_SL}(\textit b-c), separating $S$ (panel \emph{b}) and $R$ (panel \emph{c}) as well as the datasets belonging to the  H regime (top row), I regime (middle row), and T regime (bottom row) for  clarity. The time variations around the averaged value measured by the local root mean square are displayed as transparent shadings {underlying} each curve.

Figure  \ref{fig:RS_SL}(\textit b) shows that the symmetric Holmboe flows (datasets H1 and H3) have their peak $\langle S\rangle_{xyt}$ at {$z>0$; these} flows feature two trains of counter-propagating waves due to the velocity interface $\langle u \rangle_{xyt}=0$ and density interface $\langle \rho \rangle_{xyt}=0$ being nearly coincident around $z\approx 0-0.15$ (see LL22a figure 3).
By contrast, the asymmetric Holmboe flows (H2 and H4) have their peak $\langle S\rangle_{xyt}$  at {$z<0$; these} flows feature a single train of waves due to the velocity interface $\langle u \rangle_{xyt}=0$ being slightly offset from the density interface $\langle \rho \rangle_{xyt}=0$, the latter of which is around $z \approx -0.2$ (see LL22a figure 3). This peak is due to the `body' of the confined Holmboe wave structure described in \cite{Lefauve2018}, and  observed in figure~\ref{fig:RSh4I6T2T3}(\emph{a}). We also see an apparent plateau at $0<z<0.5$ in these datasets (see the `plateau' arrow), presumably due to the `head' of the confined Holmboe wave. 
With increasing levels of turbulence (I regime, middle row), this plateau in  $\langle S\rangle_{xyt}$ appears to develop into another peak (see the `peak2' arrow). Both the former and the newer peaks then tend to move closer to the edges ($z=\pm1$) of the turbulent shear layer  (T regime, bottom row). Their values $\approx 1.5-2$ is comparable to those in the H regime. A further interesting observation is that the temporal {root mean square} fluctuations of  $\langle S\rangle_{xyt}$ (width of the transparent shading) increase from the H to the I regime  but then decrease in the T regime. 
This trend {reflects the high fluctuations associated with intermittency.}

Figure  \ref{fig:RS_SL}(\textit c) shows that  $\langle R\rangle_{xyt}$ is nearly uniform in $z$ across the entire shear layer, with H1 and H3 having higher values, presumably  due to their higher $Re$ than H2 and H4 ($Re\approx 400$ \emph{vs} $200$). In the  `late' I regime (e.g. I7, I8)  and in the T regime, a broad peak in $\langle R\rangle_{xyt}$ is centered in the middle the shear layer, with {peak value that increases approximately monotonically.}  
Finally, unlike the shear, the rortex experiences equally high or even higher temporal fluctuations in the T regime (compared to the I regime), possibly caused by the breakdown and interaction of rortices.
This suggests that the emergence and increasing importance of the rortex is a fundamental aspect of turbulence dynamics and mixing, justifying our greater focus on $\bm{R}$ (vortical structures) than $\bm{S}$ (shear structures) in the remainder of this paper.

\section{Detailed morphology of rortices}

\label{sec:anatomy}
The magnitudes of $\bm{R}$ and $\bm{S}$ in the previous section provided us with {the} general trends of their spatial structures. 
This section tackles their more detailed morphology, and in particular the 3-D orientation of hairpin rortices, revealed by a comprehensive statistical analysis of our $\bm{R}$ datasets.

\subsection{Weighted conditional averaging (WCA) and orientation probability distribution functions (pdf's)}

In order to remove noise and to give stronger weight to stronger rortices, we first apply a `conditional sampling' method on the fields $\bm{R}(\bm{x},t)$ (containing between $0.3-1\times10^9$ points per dataset, depending on the spatial resolution and length of the time series). We condition these data by the following formula at all $(\bm{x},t)$:
\begin{equation}
\label{eq:conditional_threshold}
 \bm{R} = \left\{
    \begin{array}{ll}
      \bm{R}, &   \text{if} \ R/R_{rms} \ge k_{th}, \\[2pt]
      0, & \text{if} \ 0 < R/R_{rms} < k_{th},
  \end{array} \right.
\end{equation}
where $k_{th}$ is a tunable threshold level below which the data are discarded, and $R_{rms}$ is the {root mean square} of all non-zero $R$ values (before the above conditioning). Statistical processing is then performed on all non-zero $R$ after the above preconditioning. 

\begin{figure}
\centering
\includegraphics[width=0.85\textwidth]{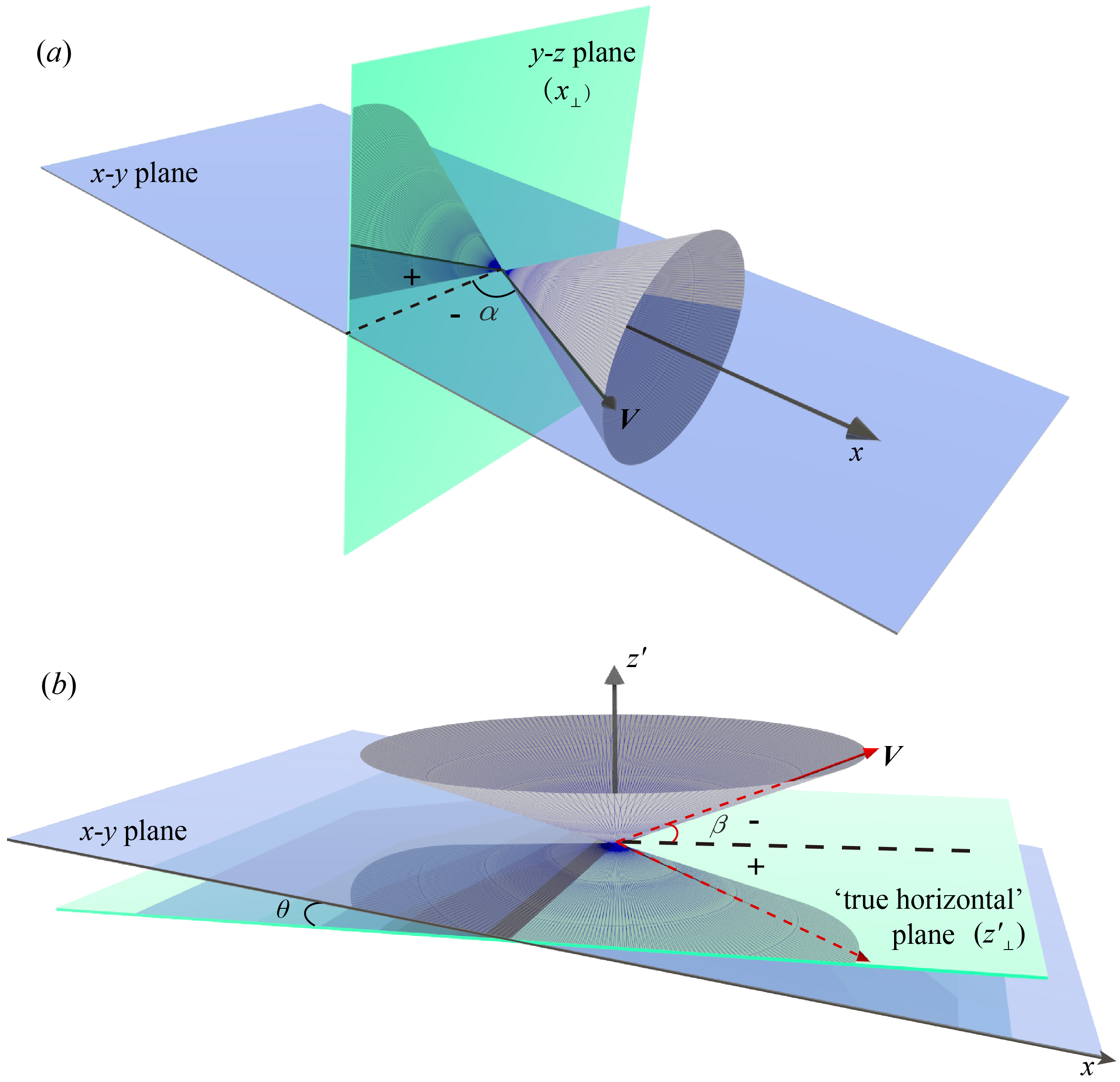}
\caption{Definition of the angles (\textit a) $\alpha = \angle(\bm{V},\bm{\hat{x}})-90$ and (\textit b) $\beta = \angle(\bm{V},\bm{\hat{z}}')-90$, where $\bm{V}$ may represent $\bm S$, $\bm R$, or $\bm{\nabla}\rho$, leading to the six angles summarised in table~\ref{Anglenotation}. The grey cones correspond to the locus of possible $\bm{V}$ for a single value of  (\textit a) $|\alpha|$ and (\textit b) $|\beta|$. The `true horizontal' plane ${z'}_\perp$ (in green in  \textit b)  is normal to the opposite direction of gravity $\bm{\hat{z}}'$, while the plane ${{x}_\perp}$  (in green in \textit a) is normal to $\bm{\hat{x}}$. Finally, $\theta$ is the tilt of the duct with respect to the horizontal direction (the convention is that $\theta>0$ indicates that the flow is forced). }
\label{fig:SchematicAngle}
\end{figure}

Next, we use `orientiation probability density functions' (pdf's) to quantify the likelihood  of the orientation of specific vectors, measured by their frequency distribution in our datasets. 
For any non-zero vector $\bm{V}$ (where $\bm{V}$ may represent $\bm{R}$, $\bm{S}$, etc) we define the angles between $\bm{V}$ and the planes ${{x}}_\perp$ and ${{z'}_\perp}$ as:
%
\begin{equation}
\alpha = \angle(\bm{V},\bm{\hat{x}})-90, \quad \beta = \angle(\bm{V},\bm{\hat{z}}')-90, \quad \bm{\hat{z}}'=\frac{-\bm{g}}{g},
\label{angleDefinition}
\end{equation}
where $\angle(\bm a,\bm b)\equiv \arccos(\bm a \cdot \bm b/(|\bm a | \, |\bm b|))\in\left[0,180\right]$ is the angle in degrees between $\bm a$ and $\bm b$. The unit vectors are defined as follows: $\bm{\hat{x}}, \bm{\hat{z}}$ are the unit vectors along the streamwise ($x$) and wall-normal ($z$) direction of the duct; ${{x}_\perp}$ indicates the plane normal to $\bm{\hat{x}}$; $\bm{\hat{z}}'=  \cos\theta \bm{\hat{z}}-\sin\theta \bm{\hat{x}}$ is the `true vertical' unit vector (in the opposite direction of gravity); and ${{z'}_\perp}$ is the `true horizontal' plane normal to the $\bm{\hat{z}}'$. 
These coordinate systems and angles (with their sign) are illustrated in figure \ref{fig:SchematicAngle} and our angle notation  is summarised in table \ref{Anglenotation}.

\begin{table}
\renewcommand\arraystretch{1.5}
\setlength{\tabcolsep}{8pt}
  \begin{center}
  \def~{\hphantom{0}}
\begin{tabular}{c c | c c | c c | c}
        $\alpha_S$&$\beta_S$&$\alpha_R$ &$\beta_R$&$\alpha_\rho$&$\beta_\rho$&$\phi$\\
   $\angle(\bm{S},x_\perp)$  & $\angle(\bm{S},z_\perp')$ &$\angle(\bm{R},x_\perp)$&$\angle(\bm{R},z_\perp')$&$\angle(\bm{\nabla}\rho,x_\perp)$ &$\angle(\bm{\nabla}\rho,z_\perp')$&$\angle(\bm{\nabla}\rho,\bm{R})$ \vspace{-0.1cm} \\ 
   $\quad \ -90^\circ$ &  $\quad \   -90^\circ$ & $\quad \  -90^\circ$ & $\quad \ -90^\circ$ & $\quad \  -90^\circ$ & $\quad \   -90^\circ$ & \\
       \multicolumn{2}{c|}{Figure \ref{fig:SGXangle}} &  \multicolumn{2}{c|}{Figures \ref{fig:turb_frac_fit_RxRg_new_log},\,\ref{fig:RxgpeakAngle},\,\ref{fig:schematic3}}& \multicolumn{2}{c|}{Figure \ref{fig:gradrho-xg}} & Figure \ref{fig:angleRgradrho}
\end{tabular}
\caption{Summary of the angles discussed in this paper (refer to definition \eqref{angleDefinition} and figure \ref{fig:SchematicAngle}). The bottom row indicate the figures in which their distributions are shown.}
\label{Anglenotation}
 \end{center}
\end{table}

Finally, to extract detailed rortex morphology from orientation pdf's, we weigh the occurrence of each value within a particular interval (histogram value)  by the local value of the `rorstrophy' (the squared rorticity) $R^2(\bm{x},t)$. This weight gives more importance to occurrences that locally coincide with high rortex values.  Practically, the averaged orientation pdf $N(i,k_{th})$ of any angle $\alpha$ or $\beta$ at a value of $i \in [ -90^\circ, 90^\circ]$ and for a given conditional threshold of $k_{th}$ is calculated by
\begin{equation}
     \begin{array}{ll}
      N(i,k_{th}) =  \dfrac{\sum_{l=1}^{n_t}\sum_{j=1}^{n_i}{R(j,l)^2}}{n_t},  & \mathrm{with} \ \   R(j,l) = 0 \ \ \mathrm{if} \ \  \dfrac{R(j,l)}{R_{rms}}< k_{th},
      \end{array}
\end{equation}
where $n_i$ is count of the occurrences when the angle under consideration belongs to the interval (bin) $i\pm\delta i$, $j$ is the index for all $(x,y,z)$ data points belonging to this interval, and $l$ is the time index sweeping through the $n_t$ `frames' (volumes) in the dataset. Note that $\int_{-90}^{90} N \, \textrm{d}i = \langle R^2 \rangle_{xyzt}$, i.e. the  area under the curve of $N$ gives the time- and volume-averaged `rorstrophy' 
satisfying the threshold $R/R_{rms} \ge k_{th}$. If $k_{th}=0$, the original rortex field and all existing rortices are considered, following  \eqref{eq:conditional_threshold}.  

The above-described weighted  (by $R^2$) and conditional (by selecting only $R/R_{rms} \ge k_{th}$) averaging will be applied to angle frequency distributions and used to study how progressively stronger rortices are aligned with respect to ${{x}_\perp}$ and ${{z'}_\perp}$. By analogy, we will also extend our weighted conditional averaging (hereafter WCA) to $\bm{S}$ (weighing by $S^2$ and conditioning by $S/S_{rms} \ge k_{th}$).

\begin{figure}
\centering
\hspace{-1cm}
\includegraphics[width=1\textwidth]{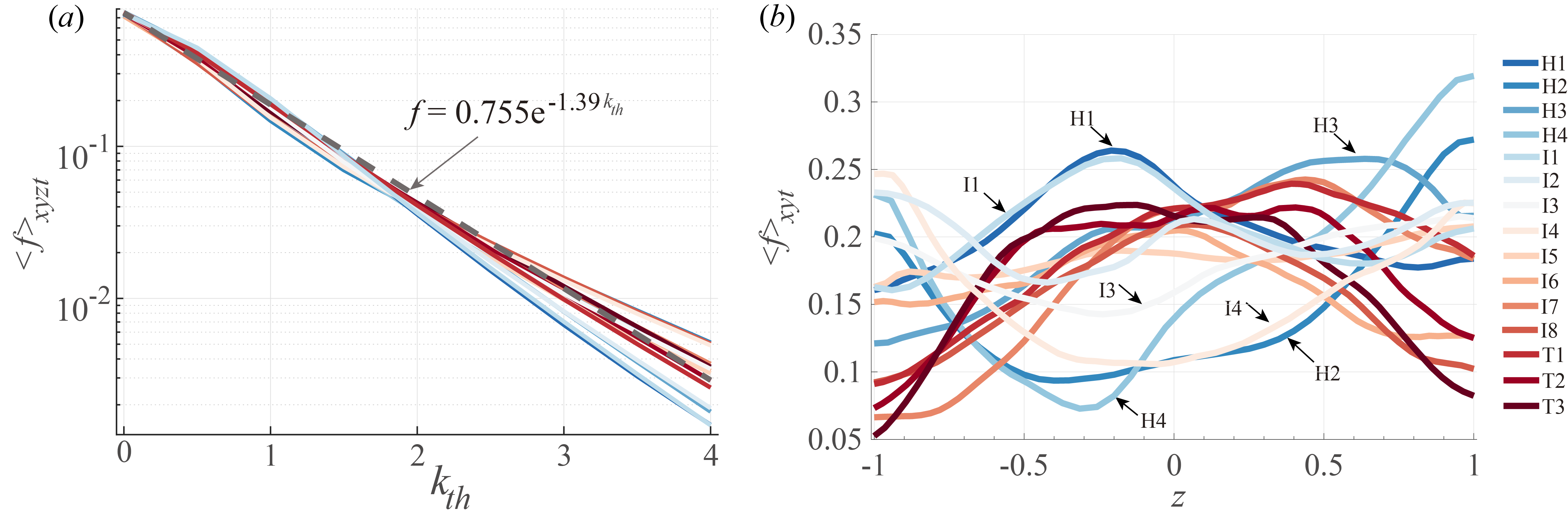}
\caption{(\textit a) Volume fraction $\langle f \rangle_{xyzt}$ of non-zero rorticity $R$ under different conditional threshold levels $k_{th}$ as defined in \eqref{eq:conditional_threshold}. Dashed line denotes the exponential fit (note the semi-log axes). (\textit b) Variation of the volume fraction $\langle f \rangle_{xyt}$ of $R$ along $\hat{z}$ for threshold $k_{th}=1$.}
\label{fig:FractionRotex_comparison}
\end{figure}

Before showing our results on the orientation of $\bm{R},\bm{S}$, it is worthwhile studying the `volume fraction of rortices' $f$ resulting from our conditional sampling method in \eqref{eq:conditional_threshold} alone without weighting. Figure \ref{fig:FractionRotex_comparison}(\textit a) shows how the global rortex volume fraction $\langle f\rangle_{xyzt} \in [0,1]$ (the time and volume-averaged ratio of points satisfying $R>k_{th}$) decreases with increasing  threshold level $k_{th}$. The semi-log axes and the exponential fit (dashed line) reveal that $\langle f\rangle_t$ decreases approximately exponentially with $k_{th}$ with decay constant $\approx 1.4$. The intercept of $0.755$ at $k_{th}=0$ means that before conditioning, approximately three quarters of the shear layer volume has non-zero rortex (i.e. the velocity gradient tensor $\bm{\nabla}\bm{v}$ has a pure rotation component).
All 15 datasets display a similar behaviour at small $k_{th}$, but their curves  spread out significantly for $k_{th}\gtrsim 2$. Increasing turbulence (curve colour transitioning from blue to red) reduces the decay rate at high $k_{th}$, confirming the intuition that turbulence leads to more frequent extreme rortex values.  Though not shown here, the volume fraction of  $R'$ (based on {the} fluctuating velocity $\bm{v}'$) is indistinguishable from that of $R$. 

Next, figure \ref{fig:FractionRotex_comparison}(\textit b) shows the vertical distribution of volume fraction $\langle f\rangle_{xyt}(z)$ (averaged in horizontal planes and time, but not $z$) at threshold $k_{th}=1$.  The result shows that rortices mainly concentrate in the middle region of the shear layer in the intermittent and turbulent datasets, which agrees with the plots of $\langle R\rangle_{xyt}(z)$ in figure \ref{fig:RS_SL}(\textit c) without  conditional thresholds. In particular T2 and T3 have a robust $\approx 20\,\%$ fraction for $|z|\lesssim 0.5$, which tapers off to only $\approx 5\,\%$ at the edges $|z| =1$, justifying our  cropping of the original datasets (see \S~\ref{sec:datasets-processing})  to restrict our attention to the $|z|\le 1$ `shear layer' containing the turbulent rortices of interest.  %
However, the H and early I regimes show slightly different tends. While datasets H1, H3 and I1 (see arrows) have a broadly similar distribution to I6-T3, datasets H2, H4 and I4 have their minimum rortex fraction near the centre of the shear layer and their maximum at the edges. This indicates that asymmetric Holmboe waves (found at high $\theta$, low $Re$) and symmetric Holmboe waves (found at low $\theta$, high $Re$), and their respective weakly intermittent turbulent counterpart (I1-I4) have inherently different rortex distribution and dynamics along $z$. This echoes the findings of LL22a (\S~6.4) that high-$\theta$, low-$Re$ turbulence is characterised by more overturning motions and less extreme shear-dominated enstrophy  than low-$\theta$, high-$Re$ turbulence.

\subsection{Inclination of shear structures}

\begin{figure}
\centering
\includegraphics[width=1\textwidth]{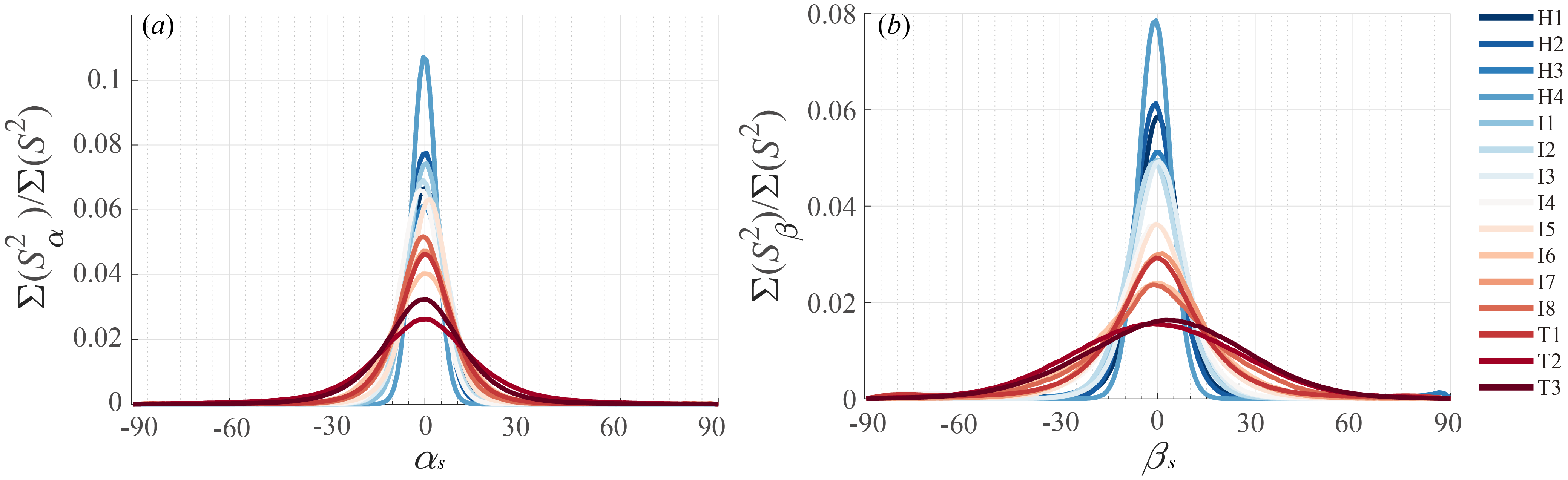}
\caption{Frequency distributions (orientation pdf's) of angles (\textit{a}) $\alpha_S$ and (\textit{b}) $\beta_S$ with threshold $S>k_{th}=2$, weighted by $S^2$. All distributions are normalised to have unit integral (like any pdf) for easier comparison between datasets.}
\label{fig:SGXangle}
\end{figure}

Although vortical structures are the focus of this study, we start with a brief study of the inclination angles of the shear vector $\bm{S}$, remembering its background role (in fact, dominant in magnitude) in shear-driven, stratified turbulence. 
Figure \ref{fig:SGXangle} shows the pdf's  of angles $\alpha_S$ and $\beta_S$ (as summarised in table~\ref{Anglenotation}). We applied our WCA method with threshold $k_{th}=2$ (thus excluding most of the modest shear associated with the mean flow $\partial_z \bar{u} = O(1)$), and weight $S^2$ (giving emphasis to large shear events).
We see that large shear events are generally perpendicular to both $\bm{\hat{x}}$ and $\bm{\hat{\bm z}}'$ and thus primarily along $\bm{\hat{y}}$, indicating the dominance of the spanwise component of vorticity, which motivated the use of $\omega_y$ in our previous study \citep{Lefauve2018}.
Increasing turbulence (blue to red curves) widens the pdf of both $\alpha_S$ and $\beta_S$, revealing increasingly 3-D shearing structures.
These results are robust for other thresholds $k_{th}$.

\subsection{Inclination of rortex structures}

\begin{figure}
\centering
\includegraphics[width=0.64\textwidth]{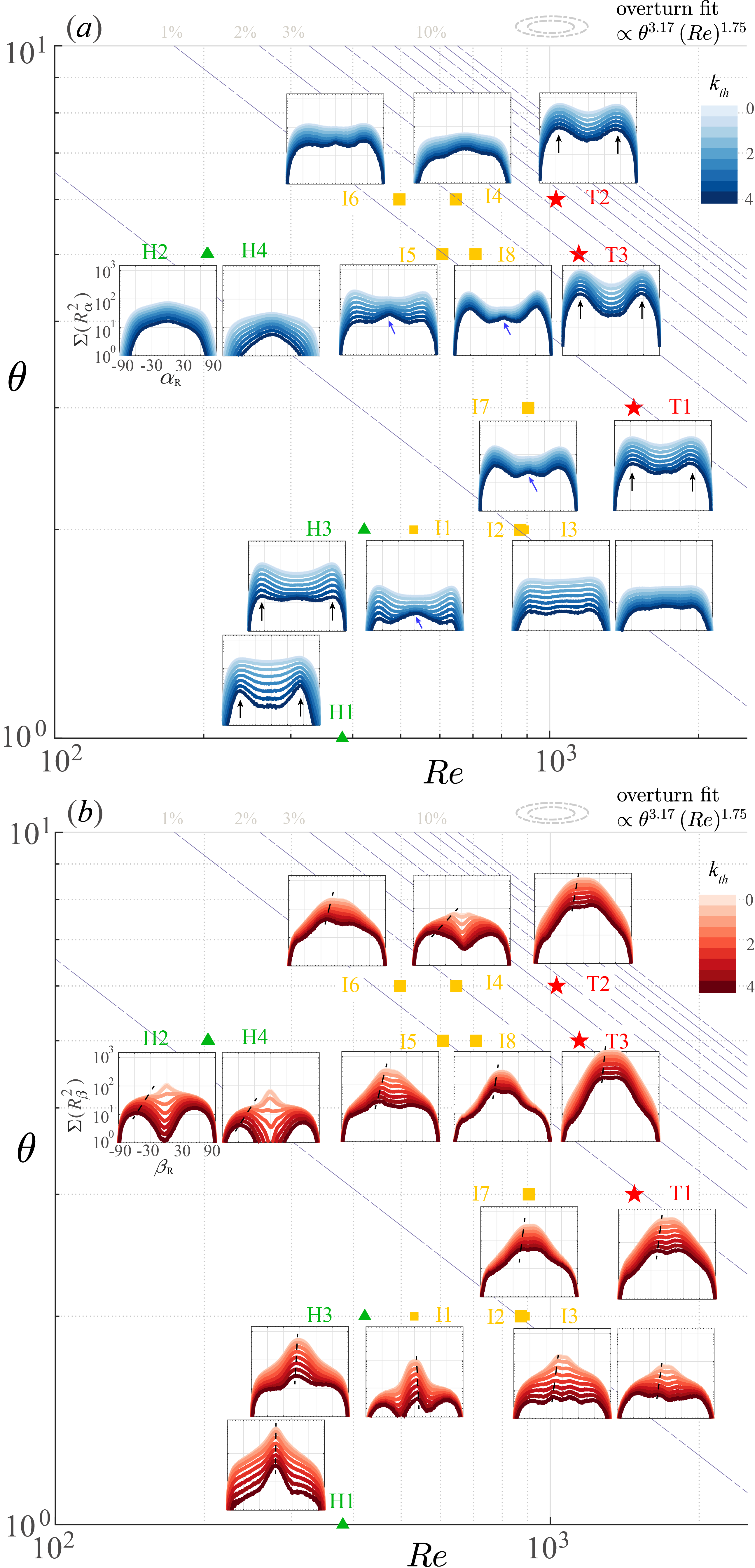}
\caption{Weighted conditionally averaged (WCA) frequency distribution (orientation pdf's) of angles (\textit a) $\alpha_R$ and (\textit b) $\beta_R$ in all 15 datasets, arranged in subpanels in the $(\theta,Re)$ plane (symbols indicate precise parameters of each dataset). We use $R^2$ weights and plot increasingly high conditional threshold levels $k_{th}\in[0:0.5:4]$ in darker shades. All panels have the same axis limits as labelled in the H2 subpanel.  The faint dashed grey lines in the background indicate the fitting of the `overturn fraction' of LL22a (their figure 8\emph{c}). Note the semi-log scale in all sub-panels.}
\label{fig:turb_frac_fit_RxRg_new_log}
\end{figure}

We now turn to a comprehensive analysis of the orientation pdf's of the vortex vector $\bm{R}$ angles, which show  different and more subtle behaviours than those of the shear.
We plot in figure~\ref{fig:turb_frac_fit_RxRg_new_log} the pdf's of $\alpha_R$ (panel~\emph{a}) and  $\beta_R$ (panel~\emph{b}). Each of the 15 datasets are plotted in separate subpanels, and arranged according to their position in the $(\theta,\Rey)$ plane, in order to draw connections between the observed vortex statistics and the two key flow parameters. Furthermore, the pdf's are weighted with $R^2$, and we use curves of increasingly dark colour (blue in panel~\emph{a} and red in panel~\emph{b}) to indicate increasingly high conditional threshold levels $k_{th}$ representative of more extreme rortices (note the semi-log scale). To facilitate comparisons of magnitudes between the 15 sub-panels, we use the same axis limits in all sub-panels and normalise all pdf's such that the integral of each over the interval $-90^\circ$ to $90^\circ$ gives the all-time and volume-averaged square norm of the conditioned $\langle R^2\rangle_{xyzt}$ (rather than 1, as in figure~\ref{fig:SGXangle}). The dashed grey diagonal lines in the background are a fit of the observed `overturn fractions' in these datasets. These were calculated in LL22a (see their \S~6) as the time- and volume-averaged fraction of the flow that experiences density overturnings ($\p_z \rho>0$), and the best fit in $(\theta, Re)$ space was shown to scale with $\theta^{3.17}\,Re^{1.75}$.  

\subsubsection{With respect to the $x_\perp$ plane}

In figure \ref{fig:turb_frac_fit_RxRg_new_log}(\textit a) we observe the following:

\begin{myenumi}

\item[\textnormal{(i)}\hspace{2.1ex}] All $\alpha_R$ pdf's are statistically symmetric around 0$^\circ$ to an excellent approximation.

\item[\textnormal{(ii)}\hspace{1.9ex}] Stronger rortices (i.e. high $k_{th}$, darker blue lines) nearly always have the same peak angle as weak rortices (i.e. low $k_{th}$, light blue lines), identified by  black arrows, except in a few datasets when an additional minor peak at $\alpha_R=0^\circ$ appears at higher $k_{th}$, identified by blue arrows.

\item[\textnormal{(iii)}\hspace{1.7ex}] A peak angle (maximum of the weighted pdf) at $\alpha_R\approx$ $\pm60^\circ$ appears in the bottom-left and top-right corners of the $(\theta,\Rey)$ plane, i.e. either at low $\theta$ and low $\Rey$ or high $\theta$ and high $\Rey$.

\item[\textnormal{(vi)}\hspace{1.5ex}] In contrast to (iii), a wider and more uniform pdf across $\alpha_R\in$  [-60$\sim$60] appears in  the upper-left and lower-right corners, i.e. either low $\theta$ and high $\Rey^s$ or at large $\theta$ and low $\Rey^s$.

\end{myenumi}

\begin{figure}
\centering
\includegraphics[width=1.03\textwidth]{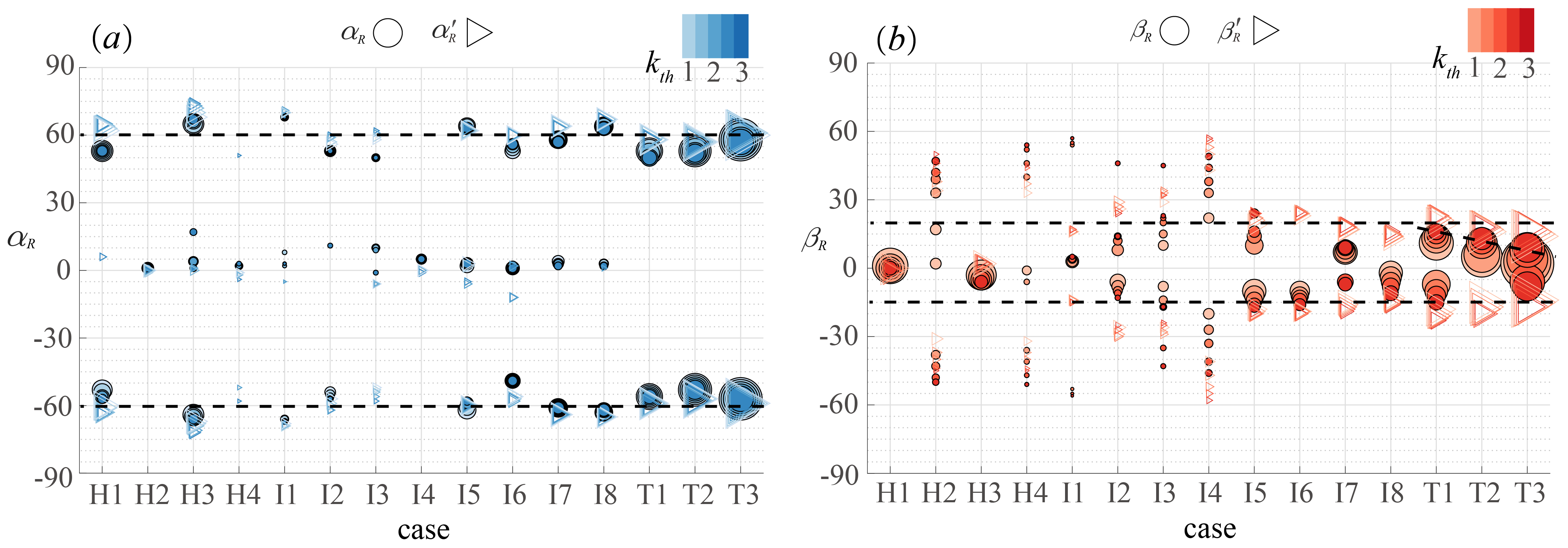}
\caption{Peak angle(s) of the  (\textit{a}) $\alpha_{R}$  and (\textit{b}) $\beta_{R}$ pdf's, automatically extracted from figure \ref{fig:turb_frac_fit_RxRg_new_log} for $k_{th}$ = 1:0.5:3 ($\iscircle$). We use $R^2$ weights and plot increasingly high conditional threshold levels $k_{th}\in[1:0.5:3]$ in darker colour (like figure~\ref{fig:turb_frac_fit_RxRg_new_log}). Open  symbols ($\vartriangleright$) denote the peak angle based on $\bm{R}’$ with the same $k_{th}$ range showing the influence of background shear. Symbol size indicate the strength of the peak angles (proportional to the square root of the ordinates $\sum R^2$ in figure~\ref{fig:turb_frac_fit_RxRg_new_log}).}
\label{fig:RxgpeakAngle}
\end{figure}

These features are clearly captured in figure \ref{fig:RxgpeakAngle}(\textit{a}), which shows the evolution of the peak values of $\alpha_R$ for values of $k_{th}$ ranging from 1 (weak rortices) to 3 (strong rortices). Larger symbols denote stronger (and thus more significant) peaks, and open symbols refer to the fluctuating rortex data $\bm{R}'$ (without $\bar{u}$), which are essentially similar to the full rortex data $\bm{R}$.

We conclude that both $Re$ and $\theta$ influence the horizontal orientation of rortices in subtle ways. In turbulent flows (T regime), both weaker and stronger rortices are primarily inclined at an angle $\alpha_R\approx\pm60^\circ$ to the ${{x}_\perp}$ plane, largely independent of $\theta$ (though it varies across a wider range than $Re$ does). However, at lower $Re$ values (H, I regimes), the evolution of $\alpha_R$ with $Re$ (at fixed values of $\theta$, i.e. along horizontal lines) varies depending on the value of $\theta$. For example, at $\theta=5^\circ$, the unimodal or uniform pdf becomes bimodal at higher $Re$  (compare the evolution H2 $\rightarrow$ H4 $\rightarrow$ I5 $\rightarrow$ I8 $\rightarrow$ T3), whereas almost the opposite happens at $\theta=2$ (evolution H3 $\rightarrow$ I1$\rightarrow$ I2 $\rightarrow$ I3). This difference again indicates that Holmboe waves at low \emph{vs} high $Re$ (here coinciding respectively with the asymmetric and symmetric type of Holmboe waves) have different properties. Rortices found in high-$Re$ (symmetric) Holmboe waves H1, H3 are more akin to those found in turbulent flows. Finally, stronger rortices can exhibit a trimodal pdf (see blue arrows) in some intermittent flows (I regime). The middle peak at $\alpha_R=0^\circ$ (perpendicular to {the} $x$-axis) suggests transverse rortices or the heads of hairpin rortices, which become less dominant under stronger turbulence. 

%

\subsubsection{With respect to the $z'_\perp$ plane (true horizontal) }

In figure \ref{fig:turb_frac_fit_RxRg_new_log}(\textit b), we observe  the following:

\begin{myenumi}

\item[\textnormal{(i)}\hspace{2.1ex}] All $\beta_R$ pdf's are also nearly statistically symmetric around $0^\circ$, even at the highest tilt angles $\theta=5^\circ$ and $6^\circ$, confirming a certain symmetry of rortices with respect to ${{z'}_\perp}$ (the true horizontal plane) rather than ${{z}_\perp}$ (the $x$-$y$ plane based on the duct coordinate system);

\item[\textnormal{(ii)}\hspace{1.8ex}] Increasingly strong rortices are inclined at increasingly steep angles to the $x-y$ plane, i.e. the peak $\beta_R$ moves away from $0^\circ$ (see the dashed trend lines in the figure), especially at low $\Rey$ and high $\theta$ (e.g. H2, H4); but this tendency diminishes somewhat (i.e. the trend line becomes more vertical) in turbulent flows (e.g. T1-T3) or at low $\theta$ and low $\Rey$ (e.g. H1, H3).

\item[\textnormal{(iii)}\hspace{1.5ex}] Both weak and strong rortices have a relatively narrow peak $\beta_R\approx \pm 10^\circ$ in the bottom-left corner of the $(\theta,Re)$ plane (H1, H3, I1). This peak widens slightly to $\beta_R\approx \pm(10$$\sim$$20)^\circ$ for the strong rortices in the top-right corner  (I8, T1, T2, T3).

\item[\textnormal{(iv)}\hspace{1.4ex}] The other pdf's (top-left and bottom-right corners of the $(\theta,Re)$ plane) tend to be more uniform or bimodal, in particular H2, H4 which have two clear peaks at $\beta_R\approx\pm(35\sim55)^\circ$.

\end{myenumi}

The evolution of the peak values for $\beta_R$ (and {$\beta_{R'}$} in open $\vartriangleright$) are plotted in figure \ref{fig:RxgpeakAngle}(\textit{b}). 
Holmboe flows are again split in two categories. In asymmetric H2 and H4 (and to some extent I1), \textLambda-rortices (without elongated tails)  are  inclined to the horizontal plane at an angle about $\pm35$$\sim$$55^\circ$, with their head inclined more steeply. In symmetric H1 and H3, rortices lie close to the horizontal plane at $0\sim\pm10^\circ$. Supplementary movie 2 shows the complete time evolution of these structures. 
In turbulent datasets, the inclination angle becomes smaller with increasing $\theta\Rey$ (see inclined dashed line). {However,} there is a {slight} difference between pdf's of $\beta_{R}$ and  $\beta_{R'}$ (open and closed symbols) in that the mean shear seems to suppress the lift-up of rortices (i.e. $\beta_R<\beta_R'$). In H2 and H4, the central peak (the maximum of the weighted pdf in figure \ref{fig:turb_frac_fit_RxRg_new_log}\textit b) $\beta_R=0^\circ$ observed for weak rortices ($k_{th}<1$) seem less due to Holmboe waves than to the mean background shear $\partial_z \bar{u}, \partial_y \bar{u}$ (which are not `pure shear'), since this peak disappears entirely when considering the pdf $\beta_{R'}$ (based on the fluctuation $\bm{R}'$, see figure \ref{fig:RxgpeakAngle}\textit{b}). However, the mean shear does not appear to `contaminate' the spanwise inclination of rortices ($\alpha_R\approx\alpha_R'$) in the more turbulent datasets (see figure \ref{fig:RxgpeakAngle}\textit{a}). 

\subsection{Inferred morphology}

\begin{figure}
\hspace{-0.8cm}
\includegraphics[width=1.05\textwidth]{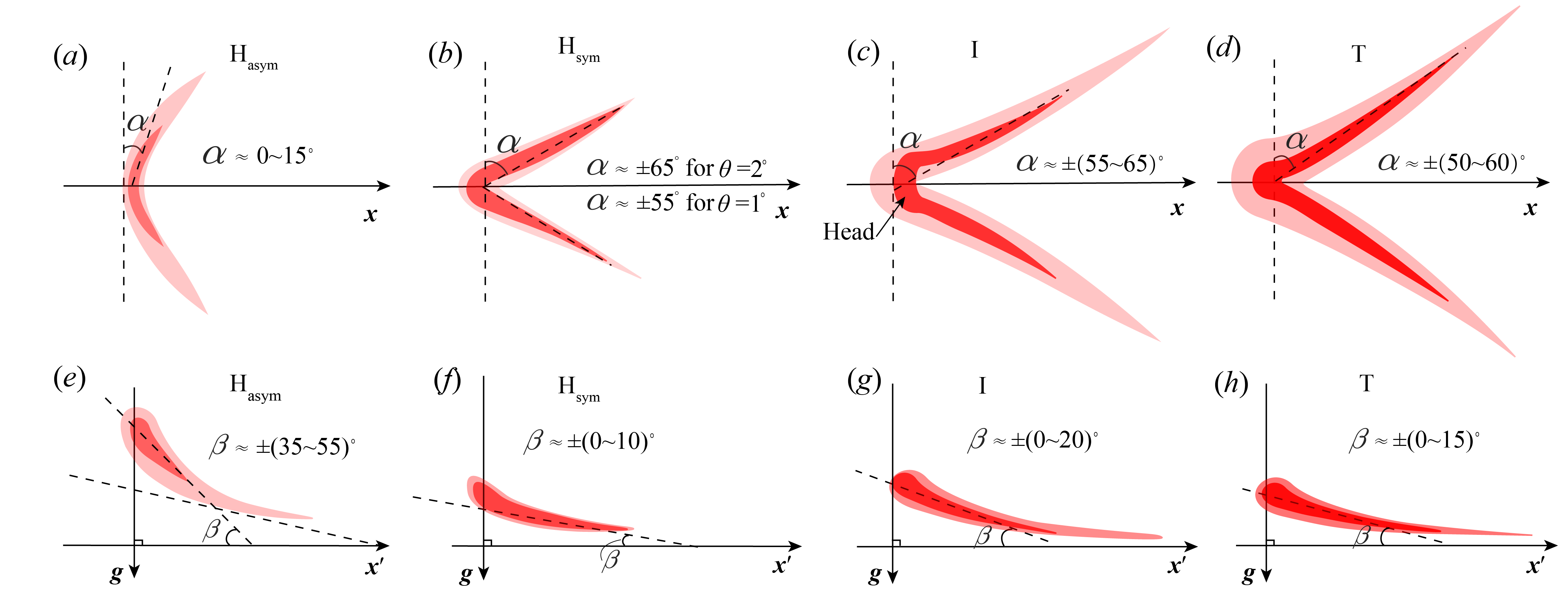}
\caption{Schematics of the evolution of rortex structures in increasingly turbulent flows. (\textit {a--d}) Top view; (\textit{e--h}) Side view. Left to right: $\rm{H_{asym}}$ and $\rm{H_{sym}}$ denote asymmetric and symmetric Holmboe regimes; I and T denote intermittently turbulent  and fully turbulent regimes, respectively.  Light red denotes weak rorticity $R$ (low $k_{th}$), dark red denotes strong $R$ (high $k_{th}$). Dashed lines indicate the typical angles of inclination found in figures~\ref{fig:turb_frac_fit_RxRg_new_log} and \ref{fig:RxgpeakAngle}.}
\label{fig:schematic3}
\end{figure}

Based on the above descriptions, we now draw {in figure \ref{fig:schematic3}} representative schematics of the morphology of rortices typical of each flow regime. 
Top views (in the $x-y$ plane) are shown in the top row (panels \textit{a-d}), while side views (in the $x'-z'$ plane, {where $x'$ is normal to $z'$}) are shown in the bottom row (panels \textit{e-h}). The strongest magnitudes $R$ are always found inside the structures and are denoted by darker shades of red (typically corresponding to $k_{th}=1$ and $2.5$). 

The angle $\alpha_R$ of the rortices in panels (\emph{b-d}) is close to that of hairpin vortices observed in unstratified turbulent boundary layers having $\alpha_R \approx 40\sim60^\circ$ \citep{Zhou1999}.
For the vertical inclination, the rortices in the T regime (panel \emph{h}) have an average $|\beta_R|\approx 11^\circ$ which agrees well with the  hairpin rortices observed in the direct numerical simulations of stratified shear layers in \cite{watanabe2019} (who reported an inclination of 14$^\circ$) and those hairpin legs observed in unstratified wall-bounded flows in \cite{Haidari_Smith1994,Zhou1999} (who reported $\approx 8^\circ$).
However, these angles are lower than the average tilt of hairpin head in {unstratified} turbulent boundary layers, which are typically around 45$^\circ$ \citep{Head1981,Zhou1999}. Interestingly,  the  `heads' of our hairpin rortices are barely more inclined than the legs, which suggests that the `lift-up' of the head is inhibited by stratification.
The legs of the symmetric Holmboe rortices (panel \emph{f})  are almost horizontal, showing a much clearer connection with intermittent and turbulent structures  than the asymmetric Holmboe ones (panel \emph{e}). 

These schematics are ideal models of symmetric $\Lambda$ and  hairpin rortices representative of the evolution of global $\alpha_R$ and $\beta_R$   statistics on either side of the turbulent transition. Instantaneous rortices in high-$Re$, turbulent, unsteady flows, are more likely to feature `broken', asymmetric hairpins, such as quasi-streamwise or cane-like rortices, as suggested by figure \ref{fig:Q} (though based on the $Q$ criterion).

\section{Interaction with density gradients}
\label{sec:interaction}

Rortices are inevitably influenced by the density (or buoyancy) field in stably-stratified flows having bulk Richardson numbers between $Ri\approx 0.15-0.55$ (in H flows) and $Ri\approx 0.1-0.2$ (in I and T flows). In this section we examine the interaction between the rortex $\bm{R}$ and density gradients $\bm{\nabla} \rho$ vectors. We first apply our weighted conditional averaging (WCA) method to the 3-D density gradient, before studying the averaged strength of interaction between rortex, shear and density gradients along the vertical. Finally, three examples, or case studies, are discussed to illustrate aspects of the complex relationship between vortical structures and mixing.

\subsection{Distribution of the density gradient and relation to the rortex}

\begin{figure}
\centering
\includegraphics[width=0.96\textwidth]{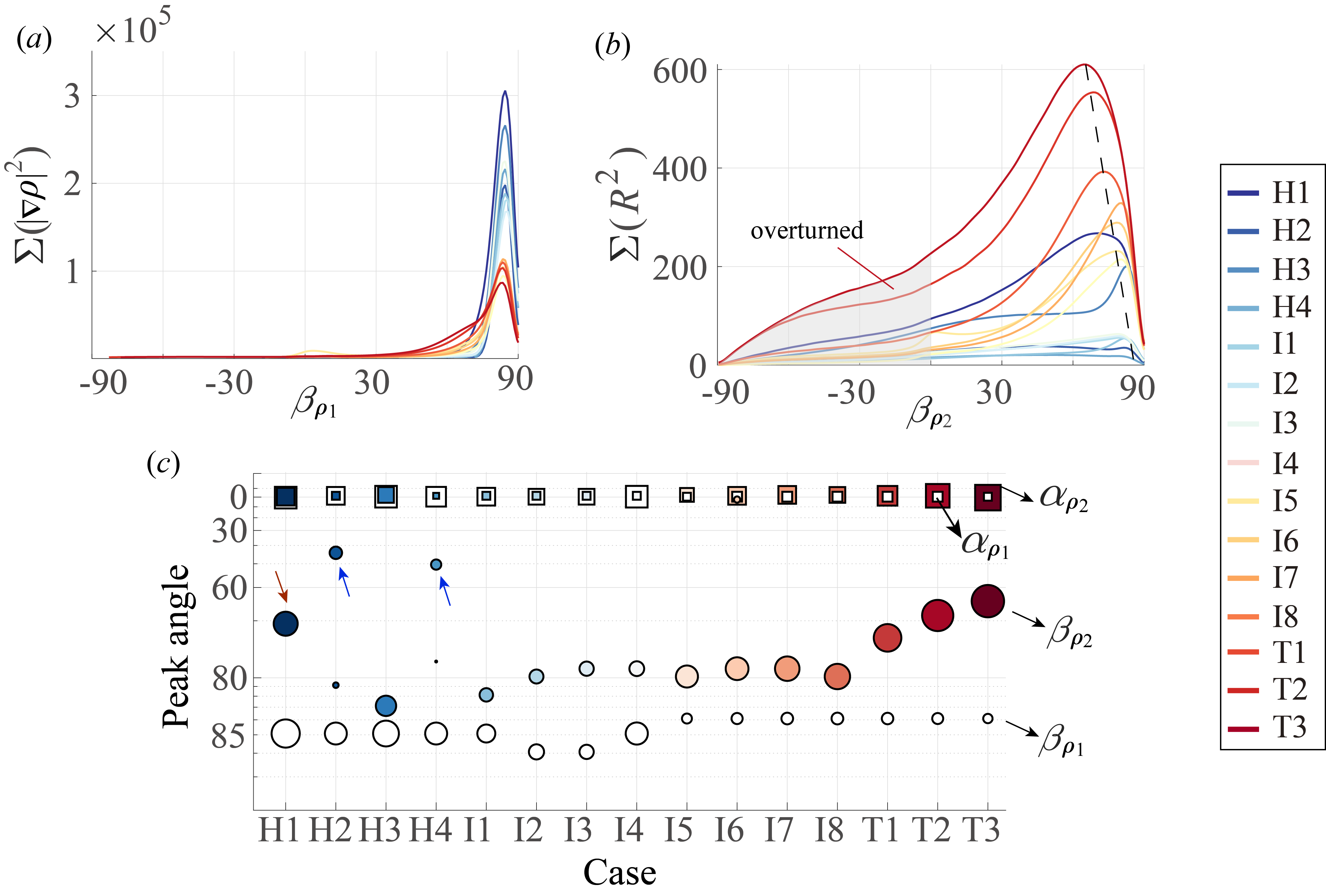}
\caption{Frequency distribution (orientation pdf's) of the WCA vertical angles of the density gradient: (\textit a) $\beta_{\rho1}$ weighted by $|\bm{\nabla}\rho|^2$ without conditional threshold; (\textit b) $\beta_{\rho2}$ weighted by $R^2$ and using a threshold of $R>k_{th}=2$. (\textit c) Peak angle (maximum of the weighted pdf) $\beta_{\rho1}$ (empty $\iscircle$)  and  $\beta_{\rho2}$  (filled $\iscircle$). We also add the horizontal angles $\alpha_{\rho1}$ with weight $|\bm{\nabla}\rho|^2$ (empty $\square$)  and  $\alpha_{\rho2}$ with weight $R^2$ (filled $\square$). Note the log vertical scale. Symbol size indicates the relative strength of the peak, i.e. the value of its ordinate in (\emph{a,b}).}
\label{fig:gradrho-xg}
\end{figure}

In figure \ref{fig:gradrho-xg} we plot the WCA pdf's of the angle between $\bm{\nabla}\rho$ and the `true horizontal' plane ${{z'}}_\perp$, i.e. $\beta_\rho = \angle(\bm{\nabla}\rho,\bm{\hat{z}}')-90$. We compare the pdf's under two different weights: in panel~(\emph{a}) with $|\bm{\nabla}\rho|^2$ to  highlight the strongest density gradients (we refer to this as $\beta_{\rho1}$); and in panel~(\emph{b}) with the squared rortex magnitude $R^2$ to highlight the orientation of density interfaces coinciding with strong rortices (we also impose the threshold $R>k_{th}=2$, and refer to this pdf as $\beta_{\rho2}$).

In panel (\textit{a}) we find, unsurprisingly for stably-stratified flows, that the strongest density gradients overwhelmingly point downwards, thus $\beta_{\rho1}>0^\circ$, with a small deviation from the perfect `true' vertical of $90-\beta_{\rho1} \approx 5^\circ$.

By contrast, in panel (\textit{b}) we observe a much broader pdf. The rorstrophy-weighted density gradients are much less vertical or downward-pointing, and a significant fraction point upward (see the grey shaded area `overturned' for $\beta_{\rho2}<0^\circ$). Density gradients that are co-located with strong rortices thus appear much more susceptible to being distorted and even overturned. The peak values of this pdf tell us about the extent of this distortion process. We observe an almost monotonic evolution from modest deviations of $90 - \beta_{\rho2} \approx 7-10^\circ$ in the H datasets to more substantial deviations of $\approx 16-25^\circ$ in the I and T datasets (see dashed trend line).  In the latter datasets, and especially in T2 and T3, the left-hand tails of the pdf's decay much more slowly, with significant overturns (we recall that the overturn fraction  was shown by light grey contours in figure \ref{fig:turb_frac_fit_RxRg_new_log}, giving a typical $\approx 3-5\,\%$ of overturned fluid in T2 and T3). These observations are robust at different $R$ conditional threshold $k_{th}$.

Figure \ref{fig:gradrho-xg}(\textit c) {summarises} the evolution of the peak angles of $\beta_{\rho2}$ (colour-filled $\iscircle$), noting in passing the difference with the evolution (or lack thereof) of $\beta_{\rho1}$ (empty $\iscircle$). The trend of a monotonically decreasing $\beta_{\rho2}$ (or increasing $90 - \beta_{\rho2}$) with $\theta Re$ is very clear. We also add the peak in horizontal angle $\alpha_\rho = \angle(\bm{\nabla}\rho,\hat{\bm{x}})-90$ ($\alpha_{\rho1}$ weighted by $|\bm{\nabla}\rho|^2$ and $\alpha_{\rho2}$ weighted by $R^2$) to show that they both have a consistent peak at 0$^\circ$.

Interpreting these results, we note that although both shear and rorticity 
increase with turbulence (as was shown in figure \ref{fig:RS_SL}), the decrease in $\beta_{\rho2}$ is presumably caused by the increasing dominance of nearly horizontal rortices having $\beta_R\approx90^\circ$ (see figure  \ref{fig:RxgpeakAngle}). These strongly-rotating hairpin `heads' (see figure  \ref{fig:schematic3})  lift up and overturn the flow around the $y$ axis. 
The symmetric Holmboe data (with {high-$Re$}, H1 and H3) are in this sense  `pre-turbulent', especially H1 (marked by a red arrow) which peaks at a 70$^\circ$.
However, the asymmetric Holmboe data (with {low-$Re$}, H2 and H4, labelled with blue arrows) are again different, since the rortex remains relatively weak compared to the shear and does not visibly influence the density gradient. The density interfaces with strongest rortices are indeed inclined at the same angle as the rortices themselves, as evidenced by the fact that $\beta_{\rho2}$ peaks at 30 $\sim40^\circ$ (labelled by blue arrows), which is very close to the peak $\beta_R$ in figure \ref{fig:RxgpeakAngle}(\textit b).  

\begin{figure}
\centering
\includegraphics[width=0.85\textwidth]{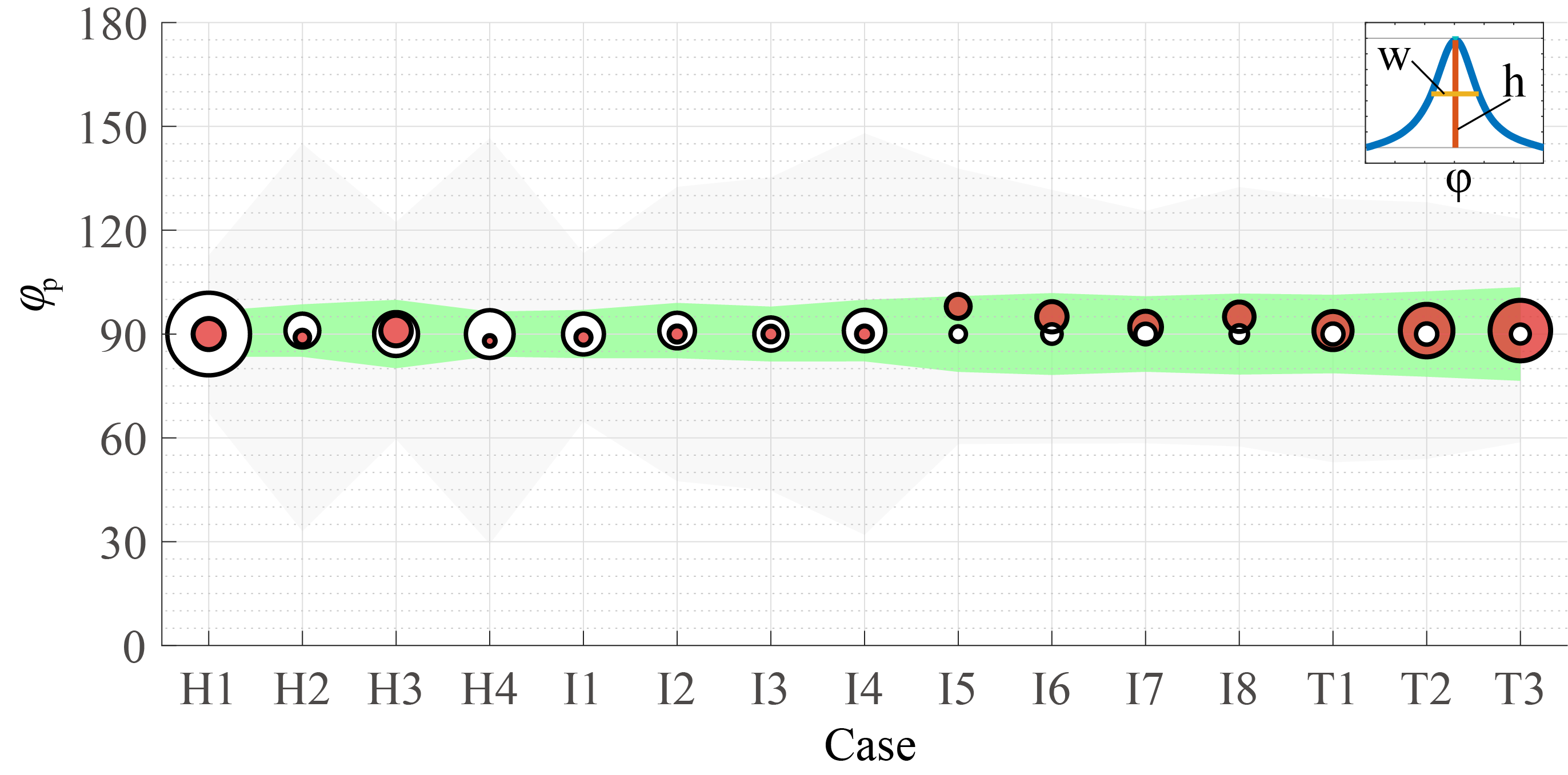}
\caption{Peak angle $\phi_p$ between  $\bm{R}$ and $\bm{\nabla}\rho$ based on WCA distributions. Symbol size  denotes the peak height $h$ (as shown schematically in the insert). Shadings denote the width $W$ of the distribution at half peak height $h/2$ (see insert). Red-filled $\iscircle$ and grey shading have weight $R^2$; white-filled $\iscircle$ and green shading have  weight  $|\bm{\nabla}\rho|^2$. All statistics  weighted by $R^2$ have conditional threshold $k_{th}=2$.   }
\label{fig:angleRgradrho}
\end{figure}

To further study the relationship between rortices and density interfaces, we define a last angle $\phi=\angle(\bm{R},\bm{\nabla}\rho)$. The evolution of its peak angle $\phi_p$, extracted from its WCA distribution, is shown in figure \ref{fig:angleRgradrho}. Symbol sizes denote the peak height, while the shading denote the width (or spread) of the distribution at half height (see top right insert).  Again we compare distributions under two different weights: $|\bm{\nabla}\rho|^2$ (white circles and green shading) and $R^2$ (red symbols and grey shading). 

First, all peak angles $\phi_{p}\approx 90^\circ$, revealing a new piece of information: rortices and density gradients are most frequently perpendicular, in all  datasets, regardless of statistical weight.

Second, distributions of $\phi$ weighted by $|\bm{\nabla}\rho|^2$   (green shading) are narrower than those weighted by $R^2$ (grey shading), having a typically spread of $\pm 10^\circ$ \emph{vs} $\pm 30^\circ$ (in some datasets even higher, e.g. H2, H4, I4). 
This observation brings an important nuance to our above conclusion: while the strongest $\bm{\nabla}\rho$ are indeed frequently perpendicular to $\bm{R}$, the strongest $\bm{R}$ are less frequently perpendicular to $\bm{\nabla}\rho$. This subtle asymmetry in the relation between $\bm{R}$ and $\bm{\nabla}\rho$ is important and understandable: we expect strong density gradients to generate rortices by baroclinic torque, but we do not expect all rortices, especially the strongest ones, to be generated by this mechanism. Our stratified layers remain dominated (driven) by shear, and most rortices can be expected to be a product of instabilities that grow by extracting energy from the mean shear, 
especially in the centre of the shear layer where the fluid is partially mixed and density gradients are weak. 

Third, although we have argued that the spread of distributions weighted by $\bm{\nabla}\rho$ (green shading) is relatively `narrow', it broadens somewhat from approximately $90\pm 5^\circ$ (H data) to $90\pm 15^\circ$ (T data) as  turbulence levels increase. This evolution shows that rortices become less perpendicular to even the strongest density gradients in increasingly turbulent flows, probably as a result of weaker stratification ($Ri \approx 0.15$ in T data) and thus of a weaker feedback of density in the momentum equation.

\subsection{Vertical profiles of their interaction}

\begin{figure}
\centering
\hspace{-0.4cm}
\includegraphics[width=1.03\textwidth]{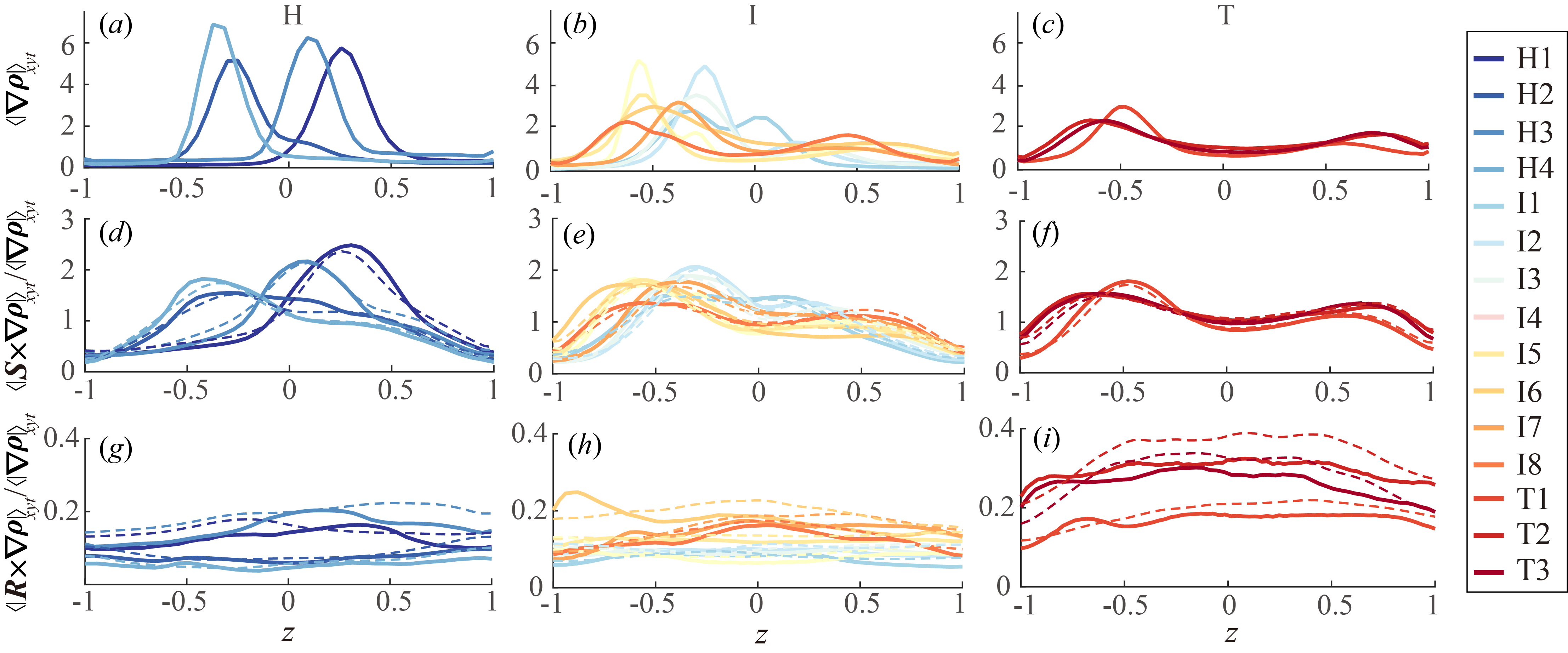}
\caption{Vertical distribution (along $z$) of the average (\textit{a-c}) $\langle|\bm{\nabla}\rho|\rangle_{xyt}$, (\textit{d-f}) $\langle|\bm S\times\bm{\nabla}\rho|\rangle_{xyt}/\langle|\bm{\nabla}\rho|\rangle_{xyt}$  and (\textit{g-i}) $\langle|\bm R\times\bm{\nabla}\rho|\rangle_{xyt}/\langle|\bm{\nabla}\rho|\rangle_{xyt}$   segregating the H, I, and T regimes in different columns. The dashed lines in figure (\textit{d-f}) and (\textit{g-i}) correspond to $\langle S\rangle_{xyt}$ and $\langle R\rangle_{xyt}$, respectively (repeating some information from  figure \ref{fig:RS_SL}\textit{b-c}).}
\label{fig:RgradrhoCross1}
\end{figure}

We now turn to the strength of $|\bm{\nabla}\rho|$ and of its interaction with $\bm{S}$ and $\bm{R}$ along  the $z$ direction. In figure \ref{fig:RgradrhoCross1}(\textit{a-c}) (top row) we show   $\langle|\bm{\nabla}\rho|\rangle_{xyt}(z)$,  segregating the H, I, and T data in different columns.  
We see that the initially sharp density interface broadens (from H $\rightarrow$ I $\rightarrow$ T)  and ends up in panel (\emph{c}) becoming partially mixed across $-0.5 \lesssim z \lesssim 0.5$, flanked by two weaker interfaces. This evolution is similar to that of the shear $\langle S \rangle_{xyt}(z)$ (previously shown in figure \ref{fig:RS_SL}\textit b).

In figure  \ref{fig:RgradrhoCross1}(\textit{d-f}) (middle row) we plot the averaged magnitude of their cross product $|\bm{S} \times \bm{\nabla}\rho|_{xyt}$ scaled by $\langle|\bm{\nabla}\rho|\rangle_{xyt}$. The profile is very close to $\langle S\rangle_{xyt}$ (shown by dashed lines), indicating that $\sin\left(\angle(\bm{S},\bm{\nabla}\rho)\right)\approx 1$, i.e. that $\bm{S}$ (primarily along $y$) is approximately perpendicular to $\bm{\nabla}\rho$ (primarily along $z$).

Figure \ref{fig:RgradrhoCross1}(\textit{g-i}) (bottom row) shows $|\bm{R} \times \bm{\nabla}\rho|$ scaled by $\langle|\bm{\nabla}\rho|\rangle_{xyt}$. The interaction of rortex and density gradient is distributed more evenly across $z$, with weak peaks that neither reflect the peak of $\langle R\rangle_{xyt}$ (in dashed lines) nor the peak  of  $\langle|\bm{\nabla}\rho|\rangle_{xyt}$ (see first row of this figure). This proves that rortices interact strongly with density gradients across the whole shear layer, rather than just at a single sharp density interface or at the two weaker interfaces on either edge of the shear layer. Although the rortex and density gradient are frequently nearly perpendicular (as we have shown in the previous figure \ref{fig:angleRgradrho}), the departure from this general tendency is substantial enough, especially in turbulent flows, that we do find any region with strongly peaked  $|\bm{R} \times \bm{\nabla}\rho|$ across the shear layer.

\subsection{Hypothesis for the role of rortices and shear on mixing}

To interpret the above observations on the different roles of shear \emph{vs} rortex on the density field, we formulate below a hypothesis for their interaction and contributions to mixing. We first recall the flow visualisation in figure \ref{fig:RSh4I6T2T3} showing strong local shear structures among {the} hairpin-like rortices that `straddle' them.  Being dominant in the vorticity field, the first contribution of the shear to mixing is to distort sharp density interfaces by shear instabilities, a process that forms vortical structures that we unequivocally identify as rortices. The role of these  rortices (weaker relative to the shear) then appears to depends on their relative strength and morphology, and on those of the density gradient. 

When rortices are weak and density gradients are stronger, such as in the Holmboe regime, rortices tend to `scour' density interfaces but can hardly destroy them, resulting in little mixing. However, when rortices are strong and density gradients weaker, such as in turbulent regime, hairpin rortices \emph{within} the shear layer (i.e. their legs) create bursting (i.e. lift-up or sweep-down events in the $z$ direction), thereby further stirring fluid within the partially-mixed layer (this is visible in the contours of velocity $v$ and $w$ in supplementary movies 2, 3, 4 and 5). 

On the other hand, the most strongly rotating parts of the hairpins \emph{at} the edges of the stratified layer (i.e. their heads) cause overturning and entrainment, thus broadening the mixing layer. Strong local shear then further stretches these newly created density gradients, accelerating small-scale molecular diffusion and ultimately achieving mixing. 

In the next section we will explore this  hypothesis and consolidate our understanding of the subtle role of rortices on mixing.

\subsection{Case studies: instantaneous snapshots of the rortex-density interactions}

\label{sec:Interaction}

The above hypothesis is based on the statistics of 15 data sets using spatial and temporal averaging, which reveals general characteristics of structures. To verify that these characteristics are indeed representative of the  actual flow phenomenology, we now study instantaneous volumetric snapshots of the rortex-density dynamics. We focus on relatively isolated rortices, inspired by the approach taken in  `kernel' studies of turbulent boundary layers for the interaction of vortical structures and the generation of near-wall bursts \citep{Haidari_Smith1994}. We shall study three datasets in turn: H4, I6 and T3, which cover the three key flow regimes of interest.

\subsubsection{H4 case}

\begin{figure}
\centering
\includegraphics[width=1.02\textwidth]{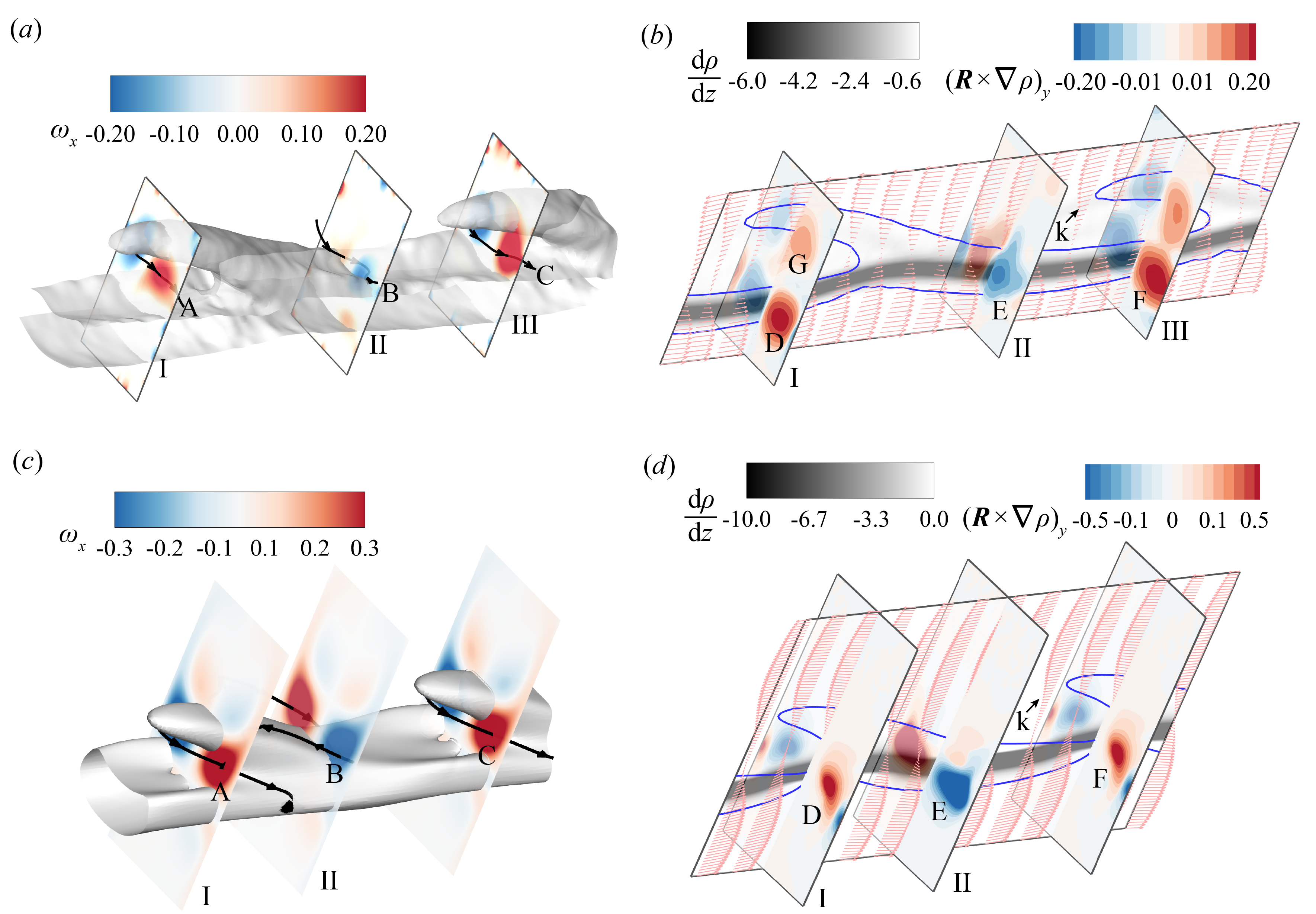}
\caption{Rortex-density interaction in asymmetric Holmboe dataset H4. Comparison between experimental (top row, \textit{a-b}) and linear stability  (bottom row, \textit{c-d}) results. (\textit a)  Streamwise vorticity ($\omega_x$, in colour) in three $y-z$ planes I, II and III, 
at time $t_n$ = 180, superimposed on an iso-surface of shear structure ($S=1.3$, in grey). (\textit b) Spanwise component of $\bm R\times\bm{\nabla}\rho$ (in colour), superimposed with the contours of $\rm{d}\rho/\rm{d}z$ (grey contour) in the $x-z$ mid-plane $y = 0$. Blue lines indicate the contour of $S=1.3$, and vectors indicate velocity in the plane. (\textit c) Same visualisation as (\emph{a}) but for the fastest-growing mode of the 3-D `confined Holmboe instability' computed in \cite{Lefauve2018}, superimposed on the iso-surface $S=2.5$. (\textit d) Same visualisation as (\emph{b}). Only the structures within the regions of $-0.8<y,z<0.8$  are shown, and the $z$ axis is stretched by a factor of 3 as in \cite{Lefauve2018}.}
\label{fig:LiushearCrossCHI}
\end{figure}

Starting with a snapshot of asymmetric Holmboe flow H4 in figure \ref{fig:LiushearCrossCHI}\textit \textit{(a-b)}, we first observe in panel (\textit{a}) that the streamwise vorticity ($\omega_x$, see colours) is mainly concentrated either under the two sides of the wave `head' or on the two sides of the wave `body' (shown by an isosurface of $S$ in grey).  The `rortex line' (black lines labelled A, B and C), equivalent to a streamline but based on the rortex vector $\bm{R}$, connects the regions of high opposite $\omega_x$ in a \textLambda ~shape. This indicates that the rortex we observed in \S~\ref{sec:Identification} likely originates from the `confined Homboe wave' of \cite{Lefauve2018} (their paper was solely based on this dataset H4).

To show the interaction between the coherent rortex and the density gradient,  we plot the $y$ component of $\bm R\times\bm{\nabla}\rho$ in $y$-$z$ planes (in colours) in panel (\textit{b}). The strongest interaction is located near the density interface where $|\bm{\nabla}\rho|$ (in grey) is largest (see the regions labelled D, E and F), recalling that here $|\bm{\nabla}\rho|$ is an order of magnitude larger than $R$ and
that the two vectors are nearly perpendicular.
High values of $(\bm R\times\bm{\nabla}\rho)_y$ are also found on either sides of the wave head (see the region labelled G), due to the high rorticity $R$.  The velocity profiles within the $x-z$ plane reveal a more inflectional  -- and thus potentially unstable -- region above the density interface, especially near the wave head (labelled k in the figure). We believe all these characteristics are important in asymmetric Holmboe waves.

In figure \ref{fig:LiushearCrossCHI}(\textit{c,d}) we show similar visualisations but for the numerical solution corresponding to the fastest growing (or most unstable) linear mode computed on the two-dimensional experimental base flow in \cite{Lefauve2018} (these data are available on the repository \cite{lefauve2018dataset}). The agreement between the observed `confined Holmboe wave' (top row) and the numerically predicted `confined Holmboe instability' (bottom row) is excellent. The only discrepancy lies in the absence of the strong interaction region that appears near the wave head in the linear solution (labelled G in panel \textit{c}). We conclude that this particular feature around the head (observed in the experimental data but absent from the linear solution) is likely caused by nonlinearities. Conversely, most other details of the rortex/density dynamics discussed above can be attributed to purely linear instability dynamics, which are significantly modulated by the spanwise confinement in the square duct geometry (for more details, see  \cite{ducimetiere_effects_2021}).

\subsubsection{I6 case}

\begin{figure}
\centering
\hspace{-0.7cm}
\includegraphics[width=1.04\textwidth]{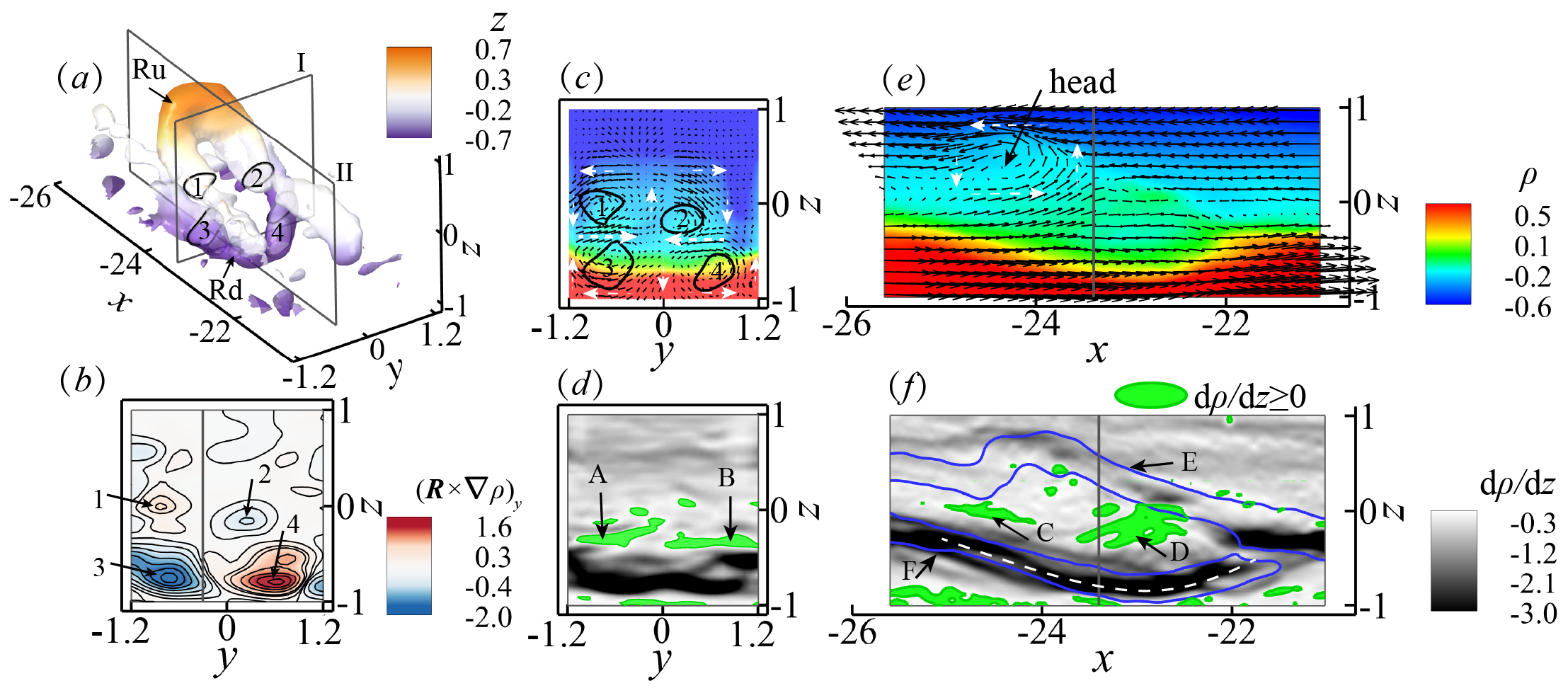}
\caption{Rortex-density interaction in intermittent dataset I6 at time  $t_n=36$. (\textit a) Isosurface of  $R = 0.6$ with colour denoting the $z$ position;  (\textit b) Spanwise  component of $\bm R\times\bm{\nabla}\rho$ in a $y-z$ (labelled I in \textit{a}); (\textit{c-d}) Density and vertical component of the density gradient in the same $y-z$ plane and in (\textit{e-f}) a $x-z$ plane (labelled II in \textit{a}). The black contour in (\textit c)  is for $R$ = 0.6; the vectors in (\textit {c-e}) show the (sub-sampled) velocity field; green contours in (\textit{d,f}) are overturned regions where $\partial_z \rho \ge0$; the blue lines in (\textit f) shows $S=1.5$. Only a subvolume (in $x$) is shown here for better visualisation.}
\label{fig:I6_t36_RgradRho_y_vec}
\end{figure}

A snapshot of I6 is shown in figure \ref{fig:I6_t36_RgradRho_y_vec}. In this intermittently turbulent flow, the life cycle (appearance and disappearance) of rortices is chaotic. Following the approach of `kernel' studies, this snapshot was selected at a time when rortices are in a relatively complete form (before their breakdown). 
Panel (\textit{a}) shows two hairpin rortices detected by the iso-surface $R=0.6$ (with its head pointing up and the other down, respectively labelled Ru and Rd) travelling in opposite directions. The rortex Ru is stronger than Rd, with the strength of their heads being $R\approx2.2$ and $R\approx0.65$, respectively. 
As time evolves (see supplementary movie 4), Ru is lifted higher up, its head finally reaches outside the shear layer while its legs are stretched through the upper layer. The overall inclination angle of Ru is $\approx30^\circ$, with its head inclined more steeply at  $\approx60^\circ$ and its leg inclined less steeply at $\approx20^\circ$. The inclination of this particular rortex is larger than the peak value (the maximum of the WCA pdf), see the I6 data in figure \ref{fig:turb_frac_fit_RxRg_new_log} and \ref{fig:RxgpeakAngle}. 
Meanwhile, the weaker rortex Rd is inclined as a whole at $13^\circ$ and its head does not lift up. The life of Rd is relatively shorter as it quickly breaks down into small eddies. 

Moving on to \ref{fig:I6_t36_RgradRho_y_vec}(\textit{b}), we see four regions of strong interaction (labelled 1 to 4) evidenced by the $y$-component of $\bm{R}\times\bm{\nabla}\rho$ in the $y-z$ plane labelled I in panel (\textit{a}). 
Again, the weaker rortex Rd seems to have a stronger interaction with the density gradient, just as in dataset H4 in the above section. 
This is due to a stronger density gradient collocated with Rd, as is clear in panels (\textit{c-d}) (showing $\rho$ and $\partial_z \rho$ in the same $y-z$ plane).

In figure \ref{fig:I6_t36_RgradRho_y_vec}(\textit{c}) we see that dense fluid (in red) under Rd is lifted up after being driven laterally outward by the rortex. On the contrary, light fluid is driven downward by the legs of Ru. The combined effect of these two rortices stirs fluid of different density (blue and red) around and make them meet in the middle region (see the white arrows in panel \textit{c}) where fluid is well mixed (cyan). This region between the two pairs of legs is where density overturns are observed (labelled A and B in figure \ref{fig:I6_t36_RgradRho_y_vec}\textit{d}). Note that the density field near Ru is better mixed than near Rd, i.e. the asymmetry between Ru and Rd carries over to the density field.

We now turn our attention to figure~\ref{fig:I6_t36_RgradRho_y_vec}(\textit{e-f}) showing the velocity vectors and density  within the $x-z$ plane `II' in panel \textit{a}. The strong head of Ru is visible and causes lift-up and sweep-down, just like the legs (see the white arrows). The overturns seen in panel (\textit{d}) are visible between the two rortices (labelled C and D in panel \textit{f}). We also note that the density interface (white dashed line) tilts towards Rd. Strong shearing structures (blue lines) are found  either at edge of the stratified shear layer (aligned with  high density gradients, see the region labelled F) or near the centre of a strong rortex (see the region labelled E). 
This upper region of high $S$ allows us to infer a relationship between the tilted rortex and the tilted region of high shear. In the H4 snapshot (figure \ref{fig:LiushearCrossCHI}), the rortex was weak,  likely created by localised shear around the Holmboe wave crest. By contrast, in this I6 snapshot, the rortex is further strengthened and stretched, and the lift up of its head also lifts up the high shear region between its two legs.

Based on the above observation, we can describe rortex Ru as a strong `stirrer' of weakly-stratified fluids, and rortex Rd as a weaker `revolving door' 
entraining denser, more strongly-stratified fluid into the mixing zone, and pulling pre-mixed fluid away from it. However this `revolving door' remains weak compared to the stratification at the interface and cannot destroy it entirely (besides, the nature of the exchange flow in the SID experiment ensures that unmixed fluid continually replaces mixed fluid, thereby sustaining such interfaces).

\subsubsection{T3 case}
\begin{figure}
\centering
\hspace{-0.2cm}
\includegraphics[width=1.0\textwidth]{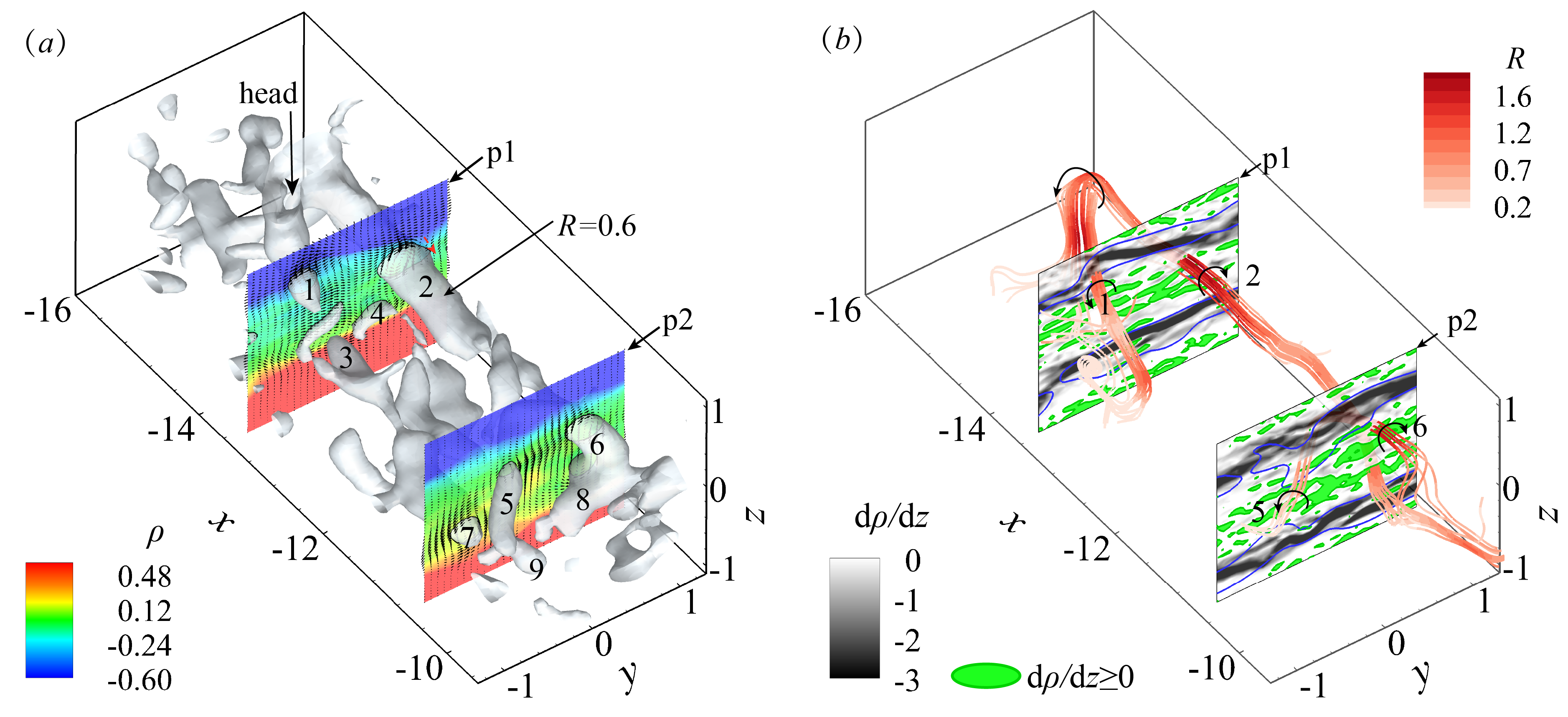}
\caption{Rortex-density interaction in turbulent dataset T3 at time $t_n=41$. (\textit a) Isourface of  $R$ = 0.6 (in grey),  with two planes (p1 and p2) showing $\rho(y,z)$. (\textit b) Vortex lines (based on $\bm{R}$) with colour denoting the strength of rortex, with the same two planes showing $\partial_z\rho(y,z)$. Green contours show density overturns; blue lines shows $S=1.5$. Only a subvolume (in $x$) is shown here for better visualisation. }
\label{fig:R_Rho_gradRho_t41}
\end{figure}

\begin{figure}
\centering
\hspace{-0.7cm}
\vspace{0.3cm}
\includegraphics[width=1.05\textwidth]{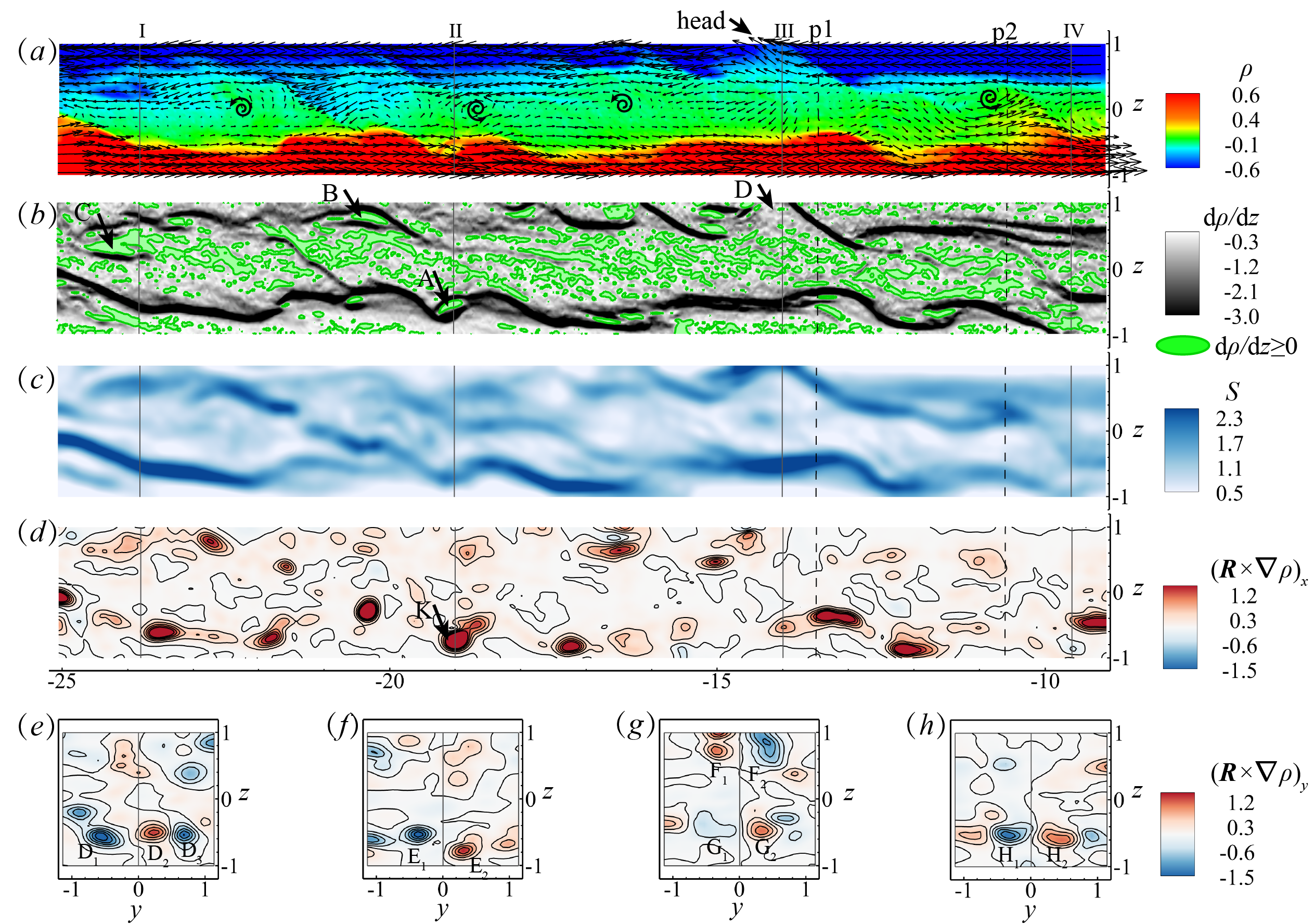}
\caption{Further rortex-density interaction in turbulent dataset T3 at time $t_n=41$ (as in figure \ref{fig:R_Rho_gradRho_t41}). (\textit {a-d}) Contours in the $y = 0$ plane of $\rho,\partial_z\rho,S$ and the  $x$-component $(\bm R\times\bm{\nabla}\rho)_x$, respectively. (\textit{e-h}) Contours in $y$-$z$ planes of the $y$-component $(\bm R\times\bm{\nabla}\rho)_y$. The solid lines in (\textit{a-d}) indicate the position of the planes (I, II, III and IV) in (\textit{e-h}), while the dashed lines indicate the position of planes p1 and p2 in figure \ref{fig:R_Rho_gradRho_t41}.} 
\label{fig:RRhograd_y_Shear_vec}
\end{figure}

In figures \ref{fig:R_Rho_gradRho_t41}-\ref{fig:RRhograd_y_Shear_vec} we select a representative snapshot in T3.
In this turbulent flow, rortices are more complex, making it more difficult to inspect isolated structures. 

Figure \ref{fig:R_Rho_gradRho_t41}(\textit a) shows the iso-surface of $R=0.6$ (grey region), $\rho$ (in colour), and $\bm{u}$ vectors in two $y-z$ planes (labelled as p1 and p2). The regions where rortices intersect these two planes are numbered 1 to 9.  Rortices 1 and 2 are the two legs of a large hairpin, whose head has been partly truncated in this figure since it protrudes outsides the shear layer $|z| \le 1$ within which our analysis is restricted. The strong rortices 1, 2, and 6 near the edges of the upper density interface move fluid laterally (around their rortex axis), thereby entraining lighter fluid (in blue) downward and neutral fluid (in green) upward.  However, due to the buoyancy restoring force, the vertical flow is less vigorous than the spanwise flow (notice the arrow length). This entrainment pattern agrees with our previous observation in I6 (figure \ref{fig:I6_t36_RgradRho_y_vec}\textit c).

Inspecting now the density gradient in figure \ref{fig:R_Rho_gradRho_t41}(\textit b), we observe that rortex 1 acts again as the typical `revolving door' described in I6, which allows for density transport across the relatively sharp upper density interface. Rortex lines,
whose colour indicate the strength $R$, intersect the $y-z$ planes in regions 1, 2, 5 and 6. The upper density interface and the corresponding high shear region (with $S=1.5$, see blue contours) between rortices 1 and 2 is visibly distorted. Between the upper and lower density interfaces (in the mixing region), rortices 3, 4, 5, and 7 stir the fluid, interact and combine, thereby increasing mixing, but the overturned region ($\p_z\rho \ge 0$, green contours) is rather spotty due to the weak stratification $\p_z\rho\approx 0$. 

Figure \ref{fig:RRhograd_y_Shear_vec} offers visualisation of the same snapshot in complementary planes (the $x-z$ plane with $y=0$ in panels \textit{a-d} and four different $y-z$ planes in panels \textit{e-h}). Based on $\rho$ (panel \textit{a}) and $\partial_z\rho$ (panel \textit{b}) the upper density interface is more irregular and slightly less stratified compared to the lower one. Velocity vectors also show that vertical motions are somewhat suppressed near the lower interface.
A strong rortex can deform and even break a nearby strong density gradient in a vertical `eruption' process across the interface (see region D where the hairpin head of the rortices 1 and 2 appears in figure \ref{fig:R_Rho_gradRho_t41}). 
Large overturning regions (filled in green, see regions A, B and C) are usually close to regions of high density gradients, sometimes forming a `sandwich' configuration.

Comparing $\partial_z \rho$ and $S$ contours in figure \ref{fig:RRhograd_y_Shear_vec}(\textit{b,c}) reveals a good correlation between them (see also the supplementary movie 5).
Although this paper mainly discusses rotational structures, it should be kept in mind that shear-driven instabilities are clearly important, and probably even dominant, in the process of turbulent production and mixing. 

Finally, we study the interaction between the rortex and the density fields, based on the $x$ and $y$ components of $\bm R\times\bm{\nabla}\rho$ in figure \ref{fig:RRhograd_y_Shear_vec}(\textit d) and (\textit{e-h}), respectively. We observe that the strongest interactions occur near the upper and lower interfaces of the stratified layer.
Peaks in the $x$ component of $\bm{R}\times \bm{\nabla} \rho$, denoted by $(\bm{R}\times \bm{\nabla} \rho)_x$,
are usually centred at the concentration of a pair of opposite values of the $y$-component $(\bm{R}\times\bm{\nabla}\rho)_y$ which again suggests a hairpin structure (see the pairs of D1-D2, E1-E2, G1-G2, F1-F2 and H1-H2 in panels \textit{e-h} and compare to the structures of panel \textit{d} at nearby $x$ locations). However, in some datasets, the opposite pair have unequal magnitude (such as G1 and G2 in panel \textit g); these rortices are then not complete hairpins, but rather cane-like or quasi-streamwise rortices, such as a rortex with a single leg connecting to a spanwise-oriented head \citep{Adrian2007}. Further comparison between panels~(\textit{b,d}) and panel (\textit{f}) shows that the `sandwich' region A is  consistent with the regions of K and E1-E2 pair, which indicates that the overturning is closely related to adjacent rortices.

\section{Synthesis and discussion}\label{sec:discussion}

Synthesising our previous statistics and case studies on rotational structures and density interfaces, we now propose a simplified model for the evolution of their morphology under increasing turbulence.

\subsection{Origin of hairpin rortices}

\begin{figure}
\centering
\includegraphics[width=0.8\textwidth]{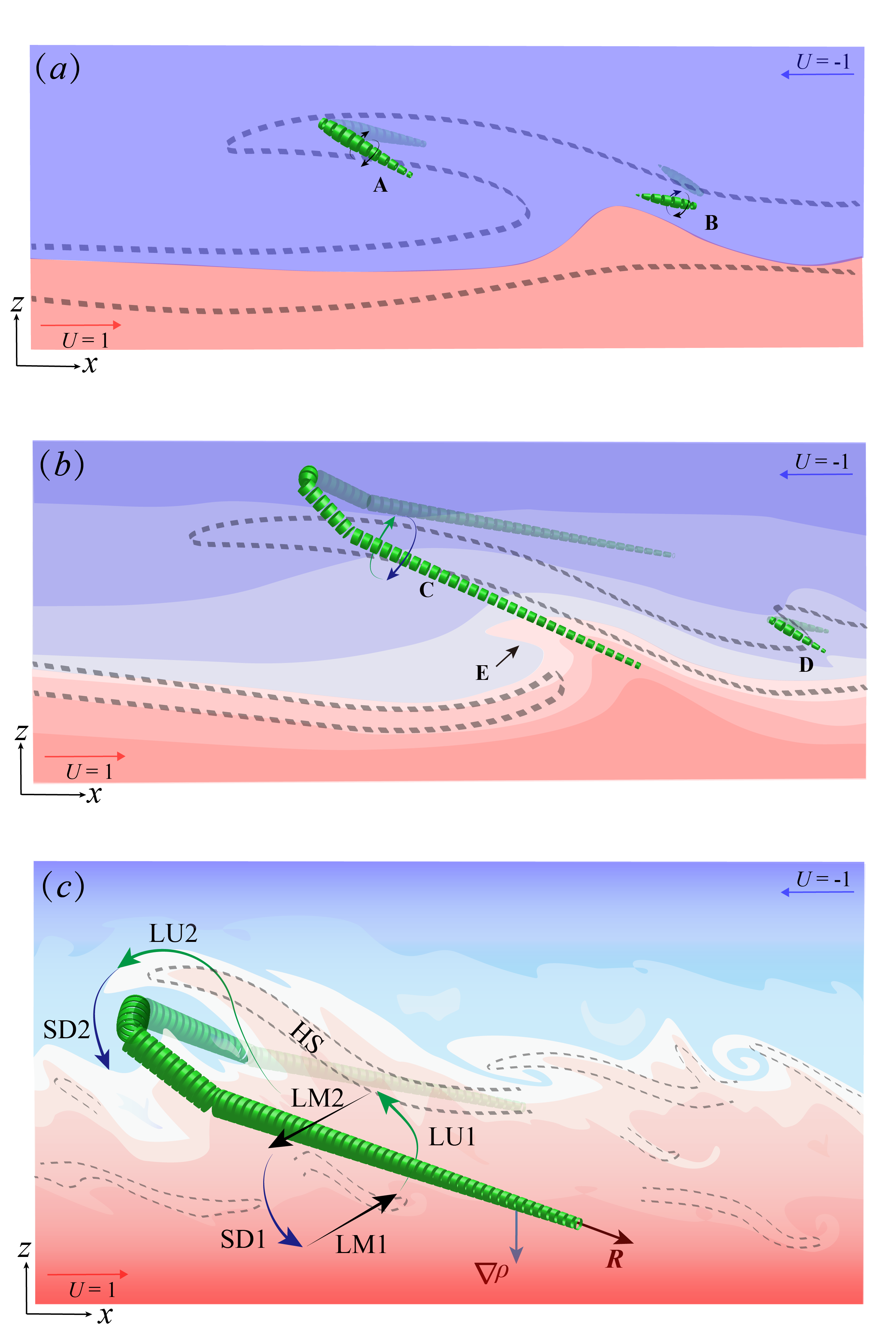}
\caption{Schematic view of (\textit a) the origin of a hairpin rortex from Holmboe wave and its evolution  in the (\textit b) intermittent regime and (\textit c) turbulent regime. Dashed lines indicate shear  structure, and green segmented tubes indicate rortices (having direction $\bm{R}$). The background colour indicates  density stratification. Abbreviations are: LU, lift up; SD, sweep down; LM, lateral movement; R, rortex vector, HS, high shear.}
\label{fig:schematicModel1}
\end{figure}

\cite{Lefauve2018} described the (asymmetric) confined Holmboe wave (CHW) in dataset H4 (here visualised in figures~\ref{fig:RSh4I6T2T3}\textit a and \ref{fig:LiushearCrossCHI}). This  long-lived  3-D wave is dominated by a shear structure having a wide `body' and a narrower `head', which are well predicted by the most unstable mode of a suitable linear stability analysis. The same study pointed out that the shear in the CHW is closely related to the coherent intensification of spanwise vorticity  due to confinement by the lateral walls of the duct, which was studied systematically in \cite{ducimetiere_effects_2021}. This vorticity intensification is reminiscent of the wave warping process \citep{Hama1963} and the vortex concentration in wall-bounded flows \citep{Smith1984}. Therefore, we speculate that the formation of rortices in SID flows is similar to the scenario of wave-induced vortex generation in a transitional boundary layer, and that \textLambda-rortices  originate from 3-D CHWs. 

A cartoon of the development of the CHW and its nearby rortices is  shown  in figure \ref{fig:schematicModel1}. 
First, the confined Holmboe instability \citep[CHI]{Lefauve2018} arises due to a resonant interaction between a  vorticity wave (on the edge of the shear layer) and a gravity wave (at the density interface). 
It then grows to a finite amplitude, intensifying and warping the background spanwise vorticity  ($\omega_y$) until it nonlinearly saturates to the observed CHW characteristic shear structure (indicated by the dashed line in figure \ref{fig:schematicModel1}\textit a).  
Subsequently, a 3-D state (with strong gradients of vorticity) likely develops according to \cite{Smyth1991} and \cite{Smyth2006}, evidenced by the nucleation of streamwise vorticity  ($\omega_x$) on either side of the wave (figure \ref{fig:LiushearCrossCHI}\emph{a-b}). 
These sites are located  under the wave `head' (or below the cusp) and above the wave body (or around the `neck'). Finally, rorticity further concentrates at these two locations,  forming a \textLambda-shaped rortex (see the embryonic rortex vectors in both the experimental and linear stability results of figure \ref{fig:LiushearCrossCHI}.
The newly generated rortices are labelled A and B in figure \ref{fig:schematicModel1}(\textit a), corresponding to the two nucleation sites labelled A and B in figure \ref{fig:LiushearCrossCHI}(\textit a,\textit c).

This above process is analogous to the wave-induced \textLambda-vortex scenario described in  boundary layers \citep{Lee2000,Jiang2020}, in that the amplification of a 3-D wave (soliton-like coherent structure) is a key initiator of a vortex, indicating that such structures may be generic to shear-driven turbulence, with and without walls. The phenomenon of Holmboe-wave-induced rortices was also suggested by the numerical simulations  of \cite{Smyth2003}. Because of the restoring force in the stratified layer, the induced embryonic \textLambda-rortex does not significantly distort the sharp density interface. The strength of the rortex is about $10~\%$  that of the shear, so the flow is still clearly shear-dominated. As pointed out by \cite{Salehipour2018} and \cite{Zhou2017}, in  strongly-stratified flows such as in the H datasets ($Ri\approx 0.1-0.6$), self-organisation occurs  through  ``scouring'' motions to keep the density interface sharp and robust.

With increasing $\theta$ and $Re$, the \textLambda-structure increases in amplitude as the rorticity is amplified by stretching, thus a more readily identifiable hairpin rortex appears (labelled C in figure \ref{fig:schematicModel1}\textit b). Its head inclines more steeply, 
due to the competition between the shear-induced stretching of the mean profile and the self-induced velocity of the \textLambda-vortex \citep{Zhou1999}.
The rortex ejects stratified fluid away from the interface, and its elongated  legs stretch into the body of the former Holmboe wave, creating transient bursts consisting of lift-up (LU) and sweep-down (SD) events, as shown in figure \ref{fig:schematicModel1}(\textit b) by the green and blue arrows, respectively. This stirs fluid above or below their original positions, creating a net buoyancy flux and production of turbulent kinetic energy from the mean shear. 
At this stage, the interface becomes more unstable, overturns more frequently,  becomes thicker (see region labelled E in the figure), and the Holmboe wave becomes shorter-lived.  Due to the presence of a strong rortex and of its induced bursting behaviours, the shear corresponding to the CHW is intensified and becomes more unstable. This localised high shear further stimulates a new rortex (e.g. the embryonic rortex labelled D in the figure), which jointly contributes to create more intense intermittent fluctuations.

Although the model shown in figure \ref{fig:schematicModel1} is inspired from  asymmetric Holmboe data, we believe the rortex generation mechanism is similar in symmetric Holmboe data. In  symmetric Holmboe waves, the typical \textLambda-rortex is more streamwise (as depicted in figure \ref{fig:schematic3}\textit b), and the streamwise stretching of the rortex legs plays a significant role during the transition to turbulence. The more horizontal the hairpin rortex is, the  easier it is for it to stir fluid up and down between its legs, generating vertical motions.

The source of the mixing in these flows is  unequivocally `internal' (following the classification of \cite{Turner1979}) in the sense that the rortex develops internally within the shear layer, which contrasts with externally introduced vortices, such as the vortex rings produced by actuating a pump \citep{Olsthoorn2015}.

The above hypothesis on hairpin rortices originating from pre-turbulent confined Holmboe waves differs from other mechanisms proposed in slightly different flows, such as the spanwise instabilities of Kelvin--Helmholtz rollers in unstable shear layers \citep{Pham2011,Pham2012} or the internal waves and quasi-linear processes in stratified plane Poiseuille flow \citep{Lloyd2022}. 

\subsection{Role of hairpin rortices in turbulent mixing}


At higher $\theta$ and $Re$, there is a stronger rortex-density interaction, more bursting and overturning because (\textit{i}) the rortices are stronger and more horizontal (i.e. they have a smaller inclination angle to the $x-y$ plane, see figure~\ref{fig:schematic3}); (\textit{ii}) a weaker stratification  ($Ri \approx 0.1-0.2$) is less able to suppress these vigorous vertical motions.

Figure \ref{fig:schematicModel1}(\textit c) shows a schematic model of rortex-density interface interaction in such conditions representative of the late intermittent regime and the turbulent regime.  Both quasi-streamwise and hairpin-like rortices are observed in the thicker, partially mixed layer bounded by two interfaces (see e.g. figure \ref{fig:RSh4I6T2T3}\textit{c,d}). Here, for simplicity, we only sketch  one hairpin rortex on the upper  interface.

The hairpin rortex consisting of a head and two legs is inclined to the true horizontal plane at a small angle (5$^\circ$ to 15$^\circ$). The legs with opposite rotation lift up  denser fluid in the middle of the rortex (see green arrow LU1 in figure \ref{fig:schematicModel1}\textit c), while on the outer side of the rortex legs, lighter fluid is swept down (see blue arrow SD1). 
This bursting cycle also enhances a lateral  exchange of momentum along the legs (see the black arrows labelled LM1 and LM2), which produces sufficiently large spanwise stress for a burst regeneration \citep{Landahl1975,Haidari_Smith1994}. Compared to vertical motions, such lateral motions in stratified turbulence are more energetic due to the absence of buoyancy restoring force in the spanwise direction.

The appearance of a distinct rortex head (or transverse rortex, along $y$), as a  manifestation of the strong concentration of spanwise rorticity, provides another bursting cycle, labelled LU2 and SD2. 
Dense fluid is lifted up (LU2) by the head, 
and overturned by the corresponding sweep-down (SD2).
Importantly, the lift-ups caused by the rortex legs and head creates a localised, inclined high-shear region slightly above the rortex (dashed line, labelled HS). Secondary instabilities of this localised high-shear produce further rortices which enhance the mixing, in a fractal-like fashion.

\begin{figure}
\centering
\includegraphics[width=1.05\textwidth]{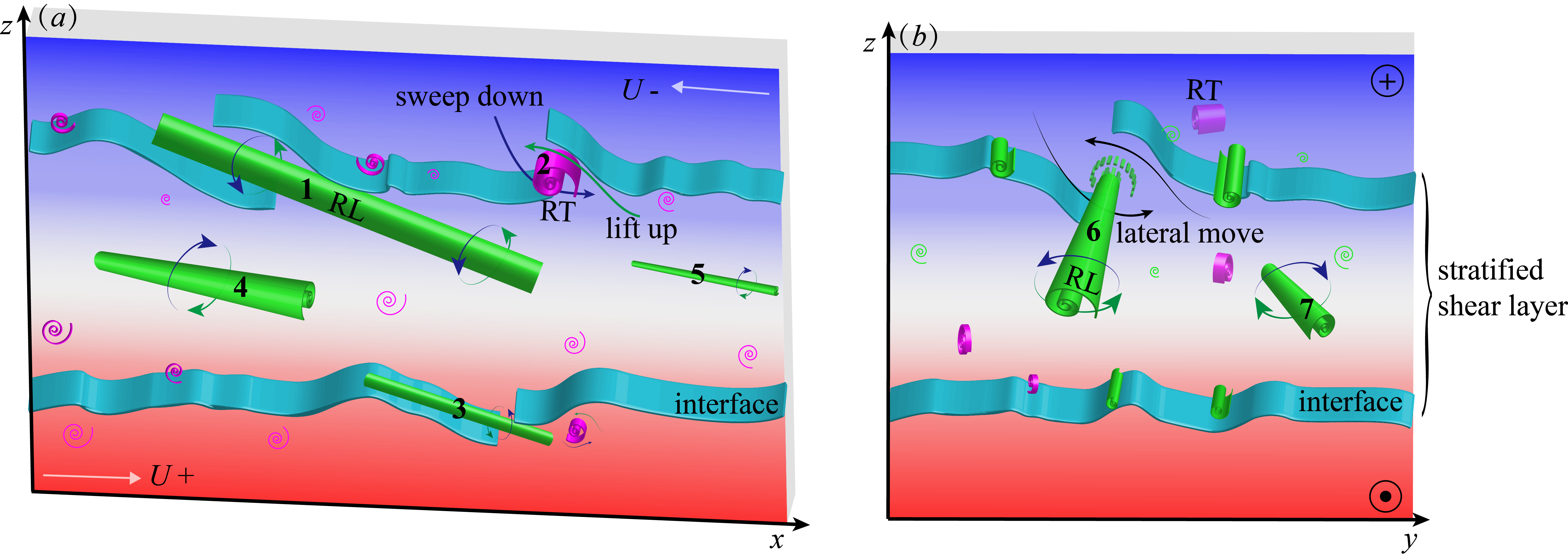}
\caption{Schematic view of turbulent rortices near a density interface and within the  partially-mixed layer. (\textit a) Side view ($x$-$z$ plane). (\textit b) Cross-sectional view ($y$-$z$ plane). RL: longitudinal rortex  (green) with $\bm{R}$ primarily along $x$; RT: transverse rortex (magenta) with $\bm{R}$ primarily along $y$.} 
\label{fig:schematicModel2}
\end{figure}

In figure \ref{fig:schematicModel2}, we further represent rortices according to their position: (\textit{i}) on the interface of the stratified layer or (\textit{ii}) within the partially-mixed layer, away from the interfaces. The green longitudinal cylinder indicates a longitudinal {rortex} (labelled RL, e.g. streamwise vortex, legs of hairpin vortex or cane-like vortex), while the magenta spiral indicates transverse {rortex} (labelled RT, e.g. spanwise vortex, hairpin head).  As explained in \S~\ref{sec:Interaction},  the rortex \emph{across} the interface acts as a `revolving door' lifting inner (pre-mixed) denser fluid away from the upper interface and entraining outer lighter fluid into the shear layer. Both LV and TV can be candidates for this role,  as shown in figure \ref{fig:schematicModel2}(\textit a). The localised breakup of the interface is closely related to this bursting cycle (see rortex 1, 2 and 3 for example). However, the rortices \emph{within} the stratified shear layer (and away from the interface, e.g. rortex 4 and 5) contribute little to this revolving door, 
being instead primarily responsible for further lateral stirring of the pre-mixed fluid, 
causing the spanwise density gradient to be further smoothed, making the interior better mixed. For rortices across the interfaces, this lateral rotation deforms the shear layer and promotes the inner-outer exchange of fluid (see rortex 6 in the cross-sectional view of figure \ref{fig:schematicModel2}\textit b).
\section{Conclusions}\label{conclusion}

In this paper, we investigated the morphology of coherent vortical structures, and more specifically `rortices', and their relation to density gradients. We adopted an empirical (data-driven) approach based on the analysis of 15 state-of-the-art experimental datasets of {increasingly} turbulent stratified shear layers obtained by exchange flow in a long inclined square duct. 
Using the standard $Q$-criterion, we first observed (figure \ref{fig:Q}) that coherent vortical structures are mainly hairpin-like (e.g. \textLambda-structures, hairpin structures, cane structures, quasi-streamwise structures) irrespective of the flow regime. After splitting vorticity into pure rotational part (rortex vector $\bm{R}$) and non-rotational part (shear vector $\bm{S}$) in figure \ref{fig:RSh4I6T2T3}, we examined their averaged magnitude $R, S$ and vertical distribution in all datasets (figure \ref{fig:RS_SL}). We found that the shear $S$ always dominates, although the rorticity $R$ increases significantly in the intermittent and turbulent regimes, and shows coherent structures reminiscent of other shear flows.  

We then studied the morphology of these coherent $\bm{R}$ structures, or `rortices', using detailed statistics (weighted conditional averaging) on the inclination angles of $\bm{R}$ with the longitudinal and vertical axes (figures \ref{fig:SchematicAngle}-\ref{fig:RxgpeakAngle}). This allowed us to draw in a key schematic (figure \ref{fig:schematic3}) the evolution of typical rortices from asymmetric (high-$\theta$ low-$Re$) and symmetric (low-$\theta$ high-$Re$) Holmboe waves (H regime), to 
{intermittent (or transitional)} flow 
(I regime), and eventually to turbulent flows (T regime). 
The two types of Holmboe waves have different rortices, the high-$Re$ symmetric ones being more similar to those found in the turbulent regime. In the 
intermittent (transitional) regime, some datasets show similarities to the asymmetric H regime, while others show similarities to the symmetric H or T regimes. Strong transverse rortices (which we attribute to `wide' hairpin heads) are most clearly observed in the I regime. 

Applying a similar statistical analysis to the density gradients and to the cross products of $\bm{R}$ or $\bm{S}$ with $\bm{\nabla}\rho$ (figures \ref{fig:gradrho-xg}-\ref{fig:RgradrhoCross1}), we found that increasingly turbulent density interfaces were increasingly steeply inclined (with respect to their true horizontal equilibrium) as a result of weaker stratification and increasing interaction with rotational structures (rortices) across the entire shear layer. By contrast, strong interaction between density gradients and shear only occurs on the edges of the partially mixed region.  We also found that while rortices can be generated baroclinically by strong density gradients, the strongest rortices were not associated with strong density gradients; on the contrary, they appear in the partially mixed region presumably as a result of shear-driven instabilities.

To complement and validate our insights based on time- and volume-average statistics, we examined in figure \ref{fig:LiushearCrossCHI}-\ref{fig:RRhograd_y_Shear_vec} the instantaneous rortex and density interface morphologies from a representative snapshot in three datasets representative of each regime. In the H regime (figure \ref{fig:LiushearCrossCHI}), the region above the density interface and near the wave head is the most unstable. The position of initial streamwise vorticity concentration agrees with the position of embryonic rortices due to a supposed secondary, nonlinear instability (which are not present in the corresponding structure predicted by a linear instability analysis). In the I regime (figure \ref{fig:I6_t36_RgradRho_y_vec}) the region between a pair of hairpin rortices (pointing up and down) is the most overturned and mixed. Strong shearing structures are located either at edge of the shear layer and aligned with region of high density gradient, or near the centre of a strong rortex. In the T regime, overturnings are frequently sandwiched by two high-density-gradient regions near the upper and lower density interfaces and flanked by nearby rortices (figures \ref{fig:R_Rho_gradRho_t41}-\ref{fig:RRhograd_y_Shear_vec}). 

To synthesise these statistical and structural results, we proposed a cartoon model for the evolution of vortical structures and density interfaces in figure \ref{fig:schematicModel1}. First, we hypothesise that \textLambda-rortices originate from the 3-D confined Holmboe waves (CHW), described in \cite{Lefauve2018}, through the formation of a  highly localised shear region, the nucleation of secondary instabilities and longitudinal roll-up. 
In turbulent flows, this rortex is strengthened and arches up, creating characteristic hairpin rortices that stir the fluid around multiple axes. Their effect on stirring (ultimately leading to mixing) is explained in figure \ref{fig:schematicModel2}. Vortices present across the upper or lower density interface act as evolving doors that drive (mixed) fluid away from the interface and entrain outer (unmixed) fluid into the mixing region. Both longitudinal rortices (e.g. hairpin legs) and transverse rortices (e.g. hairpin head) are candidates for this role. However, rortices present within the mixing region are mainly responsible for further stirring the pre-mixed fluids by lift-up/sweep-down events as well as by strong lateral movement. 

These large-scale rotational stirring motions explain the generation of a vertical buoyancy flux and the production of turbulent kinetic energy from the mean shear, which are both key energy fluxes in stratified turbulence.
At much smaller scales (not currently accessible to experimental measurements), rortices and intense shear cause turbulent kinetic energy dissipation and irreversible mixing, which are two further key energy fluxes. However, our analysis in this paper has been largely kinematic; the dynamical role of coherent rortices \textit{vs} shear in shaping the turbulent energetics -- and in particular the efficiency of mixing -- remains an open question.

\vspace{0.3cm}

\textbf{Acknowledgments}\par
The authors are grateful to Dr A. Atoufi, Dr. L. Zhu and Dr. G. Kong for valuable discussions on this research work. The authors acknowledge support from the European Research Council (ERC) under the European Union's Horizon 2020 research and innovation Grant No 742480 `Stratified Turbulence And Mixing Processes' (STAMP). A. L. is supported by a Leverhulme Trust Early Career Fellowship.

\vspace{0.3cm}

{\textbf{Declaration of interests}} \par
The authors report no conflict of interest.

\bibliographystyle{jfm}
\bibliography{Bib_short,jfm}

\clearpage

\end{document}